\documentclass[%
 reprint,
 amsmath,amssymb,
 aps,
prb,
]{revtex4-2}
\usepackage{amssymb}
\usepackage{amsthm}
\usepackage{graphicx}
\usepackage{dcolumn}
\usepackage{bm}

\begin{document}
\title{Topological phases of many-body non-Hermitian systems}
\author{Kui Cao}
\affiliation{Center for Advanced Quantum Studies, Department of Physics, Beijing Normal
University, Beijing 100875, China}
\author{Su-Peng Kou}
\thanks{Corresponding author}
\email{spkou@bnu.edu.cn}
\affiliation{Center for Advanced Quantum Studies, Department of Physics, Beijing Normal
University, Beijing 100875, China}

\begin{abstract}
We show that many-body fermionic non-Hermitian systems require two distinct sets of topological invariants to describe the topology of energy bands and quantum states respectively, with the latter yet to be explored. We identify 10 symmetry classes---determined by particle-hole, linearized time-reversal, and linearized chiral symmetries. Each class has topological invariant associated with each dimension, dictating the topology of quantum states. These findings pave the way for deeper understanding of the topological phases of many-body non-Hermitian systems.

\end{abstract}

\maketitle

\section{Introduction} Quantum Hall effect inspired developments have ushered in the topological phases paradigm for phase classification and description\;\cite{Bernevig2006, Fu2007, Moore2007, Fu2007a, Qi2008, Roy2009, Konig2007, Hasan2010, Qi2011, Read2000, Kitaev2001, Ivanov2001, Fu2008, Sato2009, Sau2010, Oreg2010, Alicea2011, Alicea2012, Sato2017}. In these phases, an underlying foundation is laid by internal symmetries. In Hermitian systems, primary internal symmetries fuse in the form of Altland-Zirnbauer (AZ) symmetry---time-reversal symmetry, particle-hole symmetry, and chiral symmetry, leading to a suite of 10 symmetry classes\;\cite{AZ}. The topological phase is classified and described by a topological invariant defined in its specific symmetry class, and all unique properties of topological phases are characterized by  topological invariant\;\cite{TCH1,TCH2,TCH3}. Changes in topological invariant always occur at points where the gap closes. Since the phase transitions of quantum states at half-filling in Hermitian systems are invariably associated with gap closing, it allows for the topological invariant to capture not only the band properties such as gapless modes but also the quantum state properties like entanglement entropy \cite{XGWEN2010,QPT2001,jcs1,jcs2}. The confines of Hermitian systems have been extended in recent years, as the inclusion of non-Hermitian effects has led to an expanded lattice of 38 symmetry classes, incorporating time-reversal symmetry, particle-hole symmetry, chiral symmetry, and three related Hermitian conjugates\;\cite{TCNH1,TCNH2,TCNH3}. In these classes, topological invariants assign distinct systems which changes tied to band gap closing.

 The effectiveness of this expanded classification framework is unquestionable in the context of energy band structures. However,  since it is hard to detect the characteristics of quantum states in many-body systems, the question remains whether it can also classify and describe the quantum state, which are specifically defined by the half-filling steady-state evolving under the dynamics of many-body non-Hermitian systems\;\cite{note2}. In the field of non-Hermitian physics, a distinct divergence from Hermitian systems has been observed. The non-Hermitian attribute, specifically the non-orthogonality of eigenstates, can induce singularities that yield phase transitions between quantum states of different phases not compulsorily occurring at the point of band gap closing\;\cite{QWG1,QWG2,QWG3,QWG4}. This implies that, in the phase diagram, the phases of quantum states may not be confined by gap closing points, unlike the topological phase diagram defined by the 38-fold classification, which is clearly limited by these points. The presence of topologically protected edge states is frequently utilized as a brief criterion for the existence of topological phases. In the single-body non-Hermitian systems, the gapless modes correspond to the topologically protected edge states, potentially leading to the belief that such edge states exist in many-body systems as well. Consequently, it is generally believed that the quantum states of these many-body systems possess same phase diagrams with band structures. However, for non-Hermitian systems, the characteristics of the quantum states of their many-body versions remarkably diverge from their single-body counterparts. Numerous instances consistent with this observation have been noted in the realm of research concerning the well-known non-Hermitian skin effect in many-body systems. Here, it is common to observe that each integral single-body wavefunction exhibits an exponential skin-like profile. Nevertheless, the manifestation of skin effect within the composite many-body wavefunction can markedly deviate from this behavior, with almost no skin effect under specific conditions \cite{ckprb,E.Lee, Shen Mu, Alsallom}. This phenomenon is interpreted as the result of non-orthogonality of the eigenstates of the non-Hermitian system, which induces overlap of particle wave functions, hence significantly enhancing the exchange force between identical particles, resulting in a significant influence on a particular particle by others. It dictates that many-body problems in non-Hermitian systems cannot be trivially distilled down to a mere issue of particle filling of their single-body counterparts \cite{ckprb}.  Addressing these concerns, our study raises two pivotal questions: Can the existing framework describe the quantum states of many-body non-Hermitian systems? If not, what alternative approach can describe the topology of quantum states? 

In this paper, we use a one-dimensional topological insulator as an example, leading to the discovery of topological phase transitions independent of gap closing, suggesting a unique property of such system: the separate of  states topology and  energy bands topology. This discrepancy also indicates that the  topology of state need necessitating distinct characterization schemes.  In light of these findings, we propose a classification approach that leads to a 10 symmetry classes base on particle-hole symmetry, linearized time-reversal symmetry and linearized chiral symmetry of Hamiltonians. Each class herein wields its distinct topological invariants associated with each dimension.  Notably, systems with significant skin effects lacking a well-defined topological invariant. Through this unveiling, we hope to advance the understanding of topological many-body non-Hermitian systems.

This paper is organized in the following way. In Section II, we taking a particular non-Hermitian topological insulator as an example, analyze phase transitions and phase diagrams for non-Hermitian topological insulators. In Section III, we introduce the topological invariants characterizing this system, necessitating two sets of topological invariants to describe the energy bands topology and the quantum states topology of the topological insulator, with the latter being previously unexplored. In Section IV, we explore the topological invariants that characterize the quantum states for general models.  Finally, in Section V, we summarize our findings.

\section{Phase diagram for non-Hermitian topological systems} 
\begin{figure}[ptb]
\includegraphics[clip,width=0.5\textwidth]{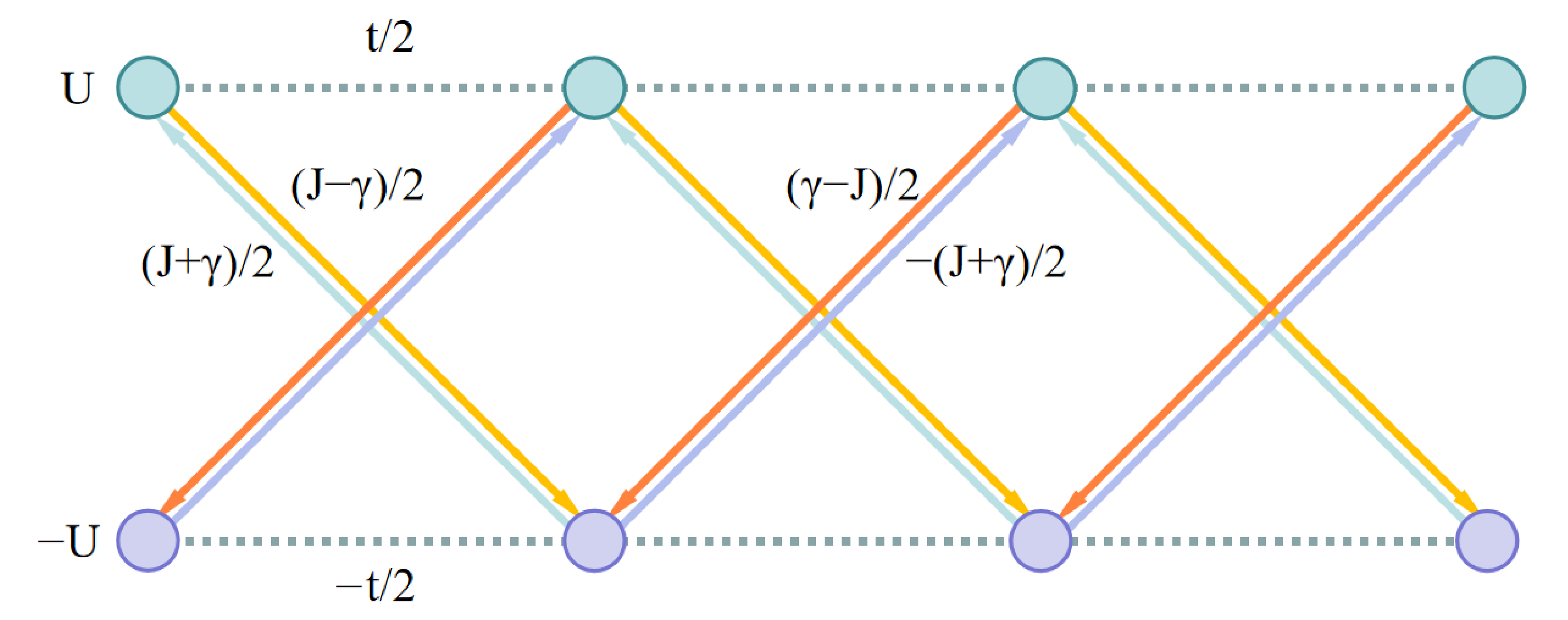}\caption{ Schematic diagram of the non-Hermitian topological insulator.} 
\label{Fig1}
\end{figure}

We begin with a half-filling 1D non-Hermitian  model   of
which  Hamiltonian is \begin{equation}\label{EQ1}\hat{H}_{\mathrm{NH}}=\sum_k \psi_{k}^\dag H_{\mathrm{NH}}(k) \psi_{k},\end{equation} where
\begin{equation}
H_{\mathrm{NH}}(k)= (U-t \cos k) \sigma_{z} +
J \sin \ k  \ \sigma_{y} - i \gamma \sin \ k \ \sigma_{x}.
\end{equation}
Here $\psi_{k}= (a_{k}, b_{k} )^{T}$,  $a_{k}$ and $b_{k}$ are the annihilation operators of fermions on sublattices $a$ and $b$ with quasi-momentum $k$, respectively. $k=\frac{2\pi n}{L},n=1,2,...,L $, here $L$ is the number of lattice.     $\sigma_{x,y,z}$ are pauli matrices. $J$, $U$, $t$, and $\gamma$ are all real numbers. We set $J, \; t>0$, consider cases where $|\gamma|\equiv \sqrt{\gamma^2}$ is less than $J$. The schematic diagram of the model is shown in Fig.\;\ref{Fig1}.

Such many-body  model can be achieved as a controlled open quantum
system $O$, composed of a Hermitian system $S$ coupling with a Markov environment $E$ 
\cite{QWG1,Yuto2018}. The Hamiltonian of the subsystem is the Hermitian part of Eq.\;(\ref{EQ1}), i.e., $\hat{H}_{\mathrm{S}}=\frac{1}{2}(\hat{H}_{\mathrm{NH}}+\hat{H}_{\mathrm{NH}}^\dag)$,
 with the set of Lindblad operator describes the coupling between sub-system $S$ and an
environment $E$:
\begin{equation}
\hat{L}_{1j}=\sqrt{2|\gamma|}(a_{j}+i \frac{\gamma}{|\gamma|}b_{j+1}), 
\end{equation}
\begin{equation}
\hat{L}_{2j}=\sqrt{2\vert \gamma\vert }(b_{j}+i\frac{\gamma}{|\gamma|}a_{j+1}).
\end{equation}
In the case of open boundary conditions, we set $j=0,1,2,...,L$ and $a_0=a_{L+1}=b_0=b_{L+1}=0$. As for the periodic boundary conditions, we set $j=1,2,...,L$ and $a_{L+1}=a_1,b_{L+1}=b_1$.
Those Lindblad operator actualized via a nonlocal Rabi coupling to auxiliary degrees of freedom paired with swift local loss  can be experimentally achieved using ultracold atoms in optical lattices proposed in Ref. \cite{ZGong2018}. 

The time evolution equation for such open system at half filling is
\begin{equation}
\frac{d}{dt}\rho=-i(\hat{H}_{\mathrm{NH}}\rho-\rho \hat{H}_{\mathrm{NH}%
}^{\dag})+\sum_{a}%
\hat{L}_{a}\rho \hat{L}_{a}^{\dag}-4|\gamma| L\rho,
\end{equation}
where $\rho$ represents the density matrix of the system, $a$ spans all possible indices of Lindblad operator $\hat{L}$.

Under full-counting measurement, the number of particles on the sub-systems $O$ is controlled to remain unchanged through post-selection (the effect of  the  post-selection on the density matrix is $\rho \rightarrow \hat{P} \rho \hat{P} / \mathrm{tr} \hat{P} \rho \hat{P}$, where $\hat{P}$ is the projection operator projecting to the subspace of specific particle number), the quantum jumping processes
represented by $\sum_{a}%
\hat{L}_{a}\rho \hat{L}_{a}^{\dag}$ are projected out, we  obtained the time evolution equation for the non-Hermitian system with 
many-body   Hamiltonian $\hat{H}_{\mathrm{NH}}$ \cite{BrodyGraefe2012,Kawabata2017}:
\begin{equation}
\label{1}
i\frac{d}{dt}\rho= \hat{H}_{\mathrm{NH}}\rho- \rho \hat{H}_{\mathrm{NH}}^{\dag} + [ \mathrm{tr} (\hat{H}_{\mathrm{NH}}^\dag-\hat{H}_{\mathrm{NH}}) \rho ] \rho,
\end{equation} 
 The first two terms correspond to the standard Liouville evolution, while the third term ensures the normalization of the density matrix.

In this model, the absence of the skin effect signifies that its characteristics can be described using the Bloch Hamiltonian $H_{\mathrm{NH}}(k)$ under both periodic boundary conditions and open boundary conditions \cite{KZhang2020,  CF2022}. The bandgap at the $k$-point is 
$$\Delta_k=2\sqrt{U^2+J^2-\gamma^2+(t^2-J^2+\gamma^2) \cos^2 k-2 Ut \cos k },$$ 
The gap $min(\Delta_k)$ closes at 
\begin{equation}U_{g\pm}=\pm t. \end{equation}


We examine the quantum states of this system. To more accurately reflect a physical setup, it's important to note that experimental systems will inevitably couple with a thermal environment. This results in an evolution equation that is equivalent to a non-Hermitian system (realized by the experimental system) coupled with a thermal environment. For the aforementioned open system $O$, which is used to realize the non-Hermitian Hamiltonian $\hat{H}_{\mathrm{NH}}$, assuming it also couples with a thermal bath $B$ with a coupling of  $\hat{H}_{ BS }\ =\sum
_{i} (\lambda^1_{i}a_{i}^{\dag}a_{i} \otimes \hat{B}^1_{i}+ \lambda^2_{i}b_{i}^{\dag}b_{i} \otimes \hat{B}^2_{i})$.  Here, $\lambda^1_{i}$,$\lambda^2_{i}$ is a
small real coupling constant and $\hat{B}^1_{i}$,$\hat{B}^2_{i}$ is an operator in thermal bath $B$. When the effect of the post-selection measurement applied on $O$ is not considered, the time evolution equation of the half-filling open system $O$ coupled with heat bath $B$ is 
\begin{equation}
\frac{d}{dt}\rho_{t}=-i(\hat{H}_{t}\rho_{t}-\rho_{t} \hat{H}_{t}^{\dag})+\sum_{a}%
(\hat{L}_{a}\otimes \hat{I}_B) \rho_{t} (\hat{L}_{a}^{\dag}\otimes \hat{I}_B) -4|\gamma| L\rho_t,
\end{equation}
here, $\rho_{t}$ is the density matrix of the whole system. The Hamiltonian  
$\hat{H}_{t}$, comprises three components:
\begin{equation}
\hat{H}_{t}=\hat{H}_{\mathrm{NH}}\otimes \hat{I}_B+\hat{I}_S\otimes \hat{H}_{B}+\hat{H}_{BS}.
\end{equation}
After applying the previously mentioned post-selection measurement, the quantum jump term $\sum_{a}
(\hat{L}_{a}\otimes \hat{I}_B) \rho_{t} (\hat{L}_{a}^{\dag}\otimes \hat{I}_B)$ will still be projected out due to the retaining of the particle number on sub-systems $O$. The system's evolution equation  is \begin{equation}
\label{1}
i\frac{d}{dt}\rho_{t}= \hat{H}_{t}\rho_{t}- \rho_{t} \hat{H}_{t}^{\dag} + [ \mathrm{tr} (\hat{H}_{t}^\dag-\hat{H}_{t}) \rho_{t} ] \rho_{t},
\end{equation}  
equivalent to coupling the non-Hermitian system $\hat{H}_{\mathrm{NH}}$ with a thermal bath.

Therefore, in this paper, to reflect physical setup, we contemplate a non-Hermitian system coupled with a thermal bath, and define its steady-state (reduced) density matrix of the non-Hermitian system as the state at temperature $T$ if it  coupled with a thermal bath at temperature $T$---this definition is consistent with the Hermitian definition. Importantly, as the temperature limits to zero, the derived density matrix reduces to the system's ground state---an aspect abundantly researched in the condensed matter physics \cite{Kcaoprr}. We considers a case of sufficiently weak coupling, such that the self-energy terms introduced by the heat bath can be neglected \cite{note}. 

%
\begin{figure}[ptb]
\includegraphics[clip,width=0.35\textwidth]{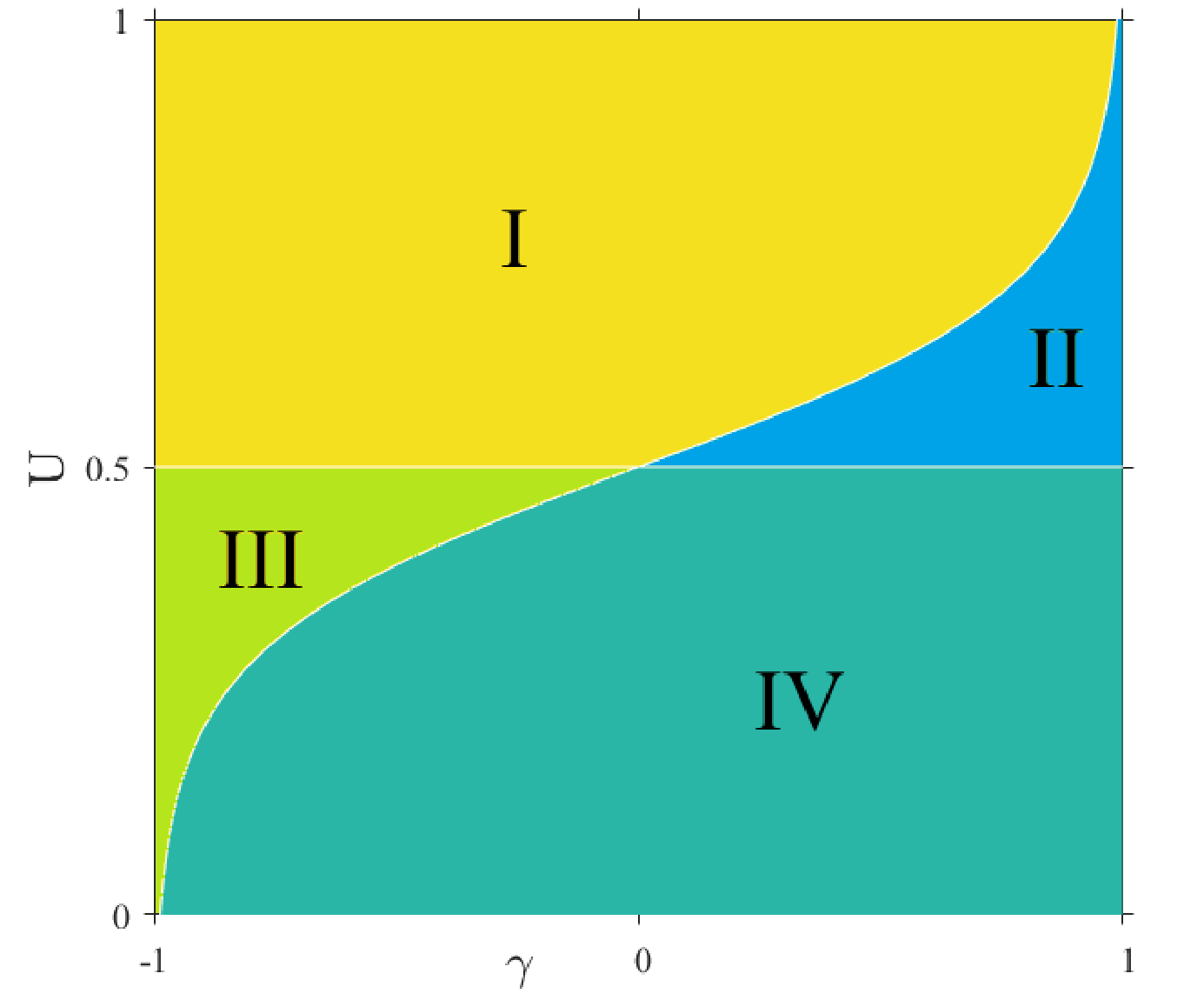}\caption{   Phase diagram for the non-Hermitian topological insulator at $T=0.2$, $t=0.5$ and $J=1$. Region I: Both quantum state and band structure are non-topological.
Region II: Topological quantum state with non-topological band structure.
Region III: Non-topological quantum state with topological band structure.
Region IV: Both quantum state and band structure are topological.} 
\label{Fig2}
\end{figure}

We obtain  the unnormalized density matrix (it generates the system's authentic density matrix upon normalization) at inverse temperature $\beta \equiv \frac{1}{T}$ as (see details in Appendix A) 
\begin{equation}
\rho= e^{ - \beta \hat{H}_{\mathrm{NH}}}  e^{\frac{1}{2}\ln  \frac{J+\gamma}{J -\gamma}  \sum_i   \psi^\dag_{i}   \sigma_z \psi_{i} } .
\end{equation}
Here, $\psi_{i}= (a_{i}, b_{i} )^{T}$,  $a_{i}$ and $b_{i}$ denote the fermion annihilation operators on the $i$-th sublattices $a$ and $b$. All operators are confined within the Fock subspace characterized by a particle number of $L$.  This formula holds for both periodic boundary conditions and open boundary conditions. In view of the fact that the density matrix of the system retains a Gaussian profile, we are in a position to define an effective Hamiltonian of quadratic form, represented as \begin{equation}\hat{H}_\mathrm{eff}\equiv-\ln \rho.\end{equation}
The Bloch Hamiltonian corresponding  to $\hat{H}_\mathrm{eff}$  emerges as
\begin{equation}
\label{EQ7}
H_\mathrm{eff}(k) =\frac{W(k)}{\sqrt{A_{y}(k)^{2}+A_{z}(k)^{2}}}\left[
A_{y}(k)  \sigma_{y}+A_{z}(k)  \sigma_{z}\right]  \nonumber,
\end{equation}
  with
\begin{align}
W(k)  &  = \cosh^{-1} \{\frac{1}{\sqrt{J^2-\gamma^2}}[J   \cosh \left(  \beta \Delta_k/2\right) \nonumber \\
&  +2\gamma (t \cos k -U)   \sinh \left(
 \beta \Delta_k/2 \right)  \Delta_k^{-1}] \} ,\nonumber \\
A_{y}(k)  &  =  \sin k  \frac{ J^2-\gamma^2 }{J} ,\nonumber \\
A_{z}(k)  &  =U-t \cos k  -  \frac{\gamma}{2J} \frac{ \Delta_k}  
{\tanh \left(   \beta \Delta_k/2 \right)}.
\end{align}

$H_\mathrm{eff}$ captures the information of the quantum state in which the non-Hermitian system resides. Specific,  the quantum state of the non-Hermitian system is mapped to the quantum state of a Hermitian system with Hamiltonian $H_\mathrm{eff}$ at unit temperature.  The closing of the band gap in $H_\mathrm{eff}$ is recognized as the phase transition point of the quantum state. For a Hermitian system, the effective Hamiltonian is proportional to the system's Hamiltonian, thus yielding the same gap closing points. Interestingly, in non-Hermitian case, the closing of the band gap in $H_\mathrm{eff}$ did not align with the closure of the system gap $U_{g\pm}$.  It occurs at
 \begin{equation}  
U_{c \pm}
=\frac{T}{2} \ln \frac{J+\gamma}{J-\gamma} \pm t. 
\end{equation}
This implying a phase transition  has occurred in the system without   gap closing.

  The  gap closing points $U_g$ and phase transition critical points $U_c$ divide the phase diagram into four regions, as shown in Fig.\;\ref{Fig2}. To understand the significance of the regions in the phase diagram, we brief analyze the implications of the two lines. 

\begin{figure}[ptb]
\includegraphics[clip,width=0.5\textwidth]{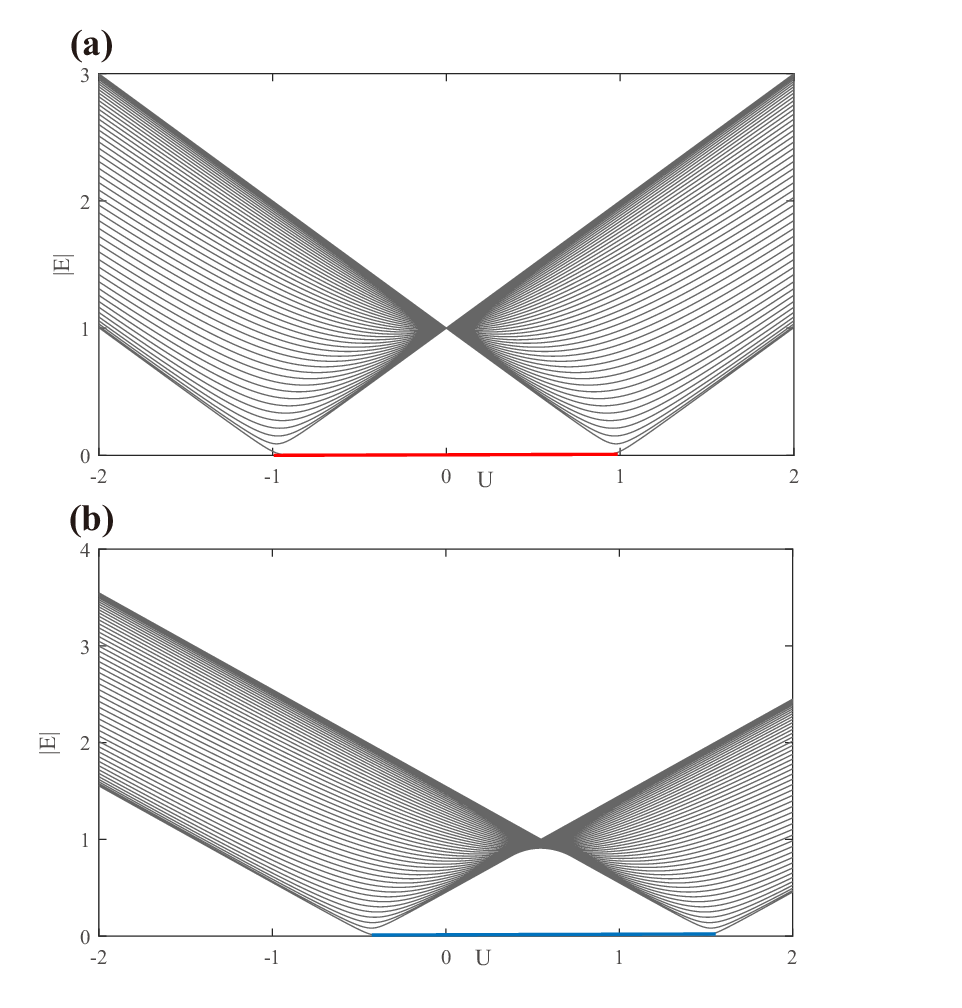}\caption{ (a) Numerical spectra for the non-Hermitian topological insulator with 50 lattice cells ($L=50$) under open boundary conditions, featuring the zero-mode line depicted in red. This line denotes a twofold degenerate state, with any inconspicuous splitting being neglected.
(b) Numerical spectra of the corresponding effective model for the non-Hermitian topological insulator, also with 50 lattice cells ($L=50$) and subjected to open boundary conditions. Here, the zero-mode line is illustrated in blue, indicating a twofold degenerate state while disregarding any negligible splitting. For these calculations, we set $T=1$, $t=1$, $J=1$ and $\gamma=0.5$.} 
\label{Fig3}
\end{figure}

Firstly, we analyze the physical significance of the band closing point $U_g$. We note that two zero modes appear among the two band closing points  $U_{g\pm}$. See Fig.\;\ref{Fig3}(a). Therefore, the critical point $U_{g\pm}$ is the topological phase transition point for the band  structure of the non-Hermitian system; when $ U_{g-}<U <U_{g+} $, the band structure of the system resides in a topological phase, whereas for $U>U_{g+}$ or $U<U_{g-}$, the band structure transitions into a trivial phase.

Secondly, we analyze the physical significance of the phase transition critical points $U_c$. We note that for the effective model which characterizes the quantum states of the system, two zero modes appear among the two band closing points  $U_{c\pm}$, see Fig.\;\ref{Fig3}(b).   Therefore, the critical point $U_{c\pm}$ is the topological phase transition point for the state of the non-Hermitian system;
when $ U_{c-}<U <U_{c+} $, the quantum states   of the system resides in a topological phase, whereas for $U>U_{c+}$ or $U<U_{c-}$, the quantum states  transitions into a trivial phase.  

\begin{figure}[ptb]
\includegraphics[clip,width=0.5\textwidth]{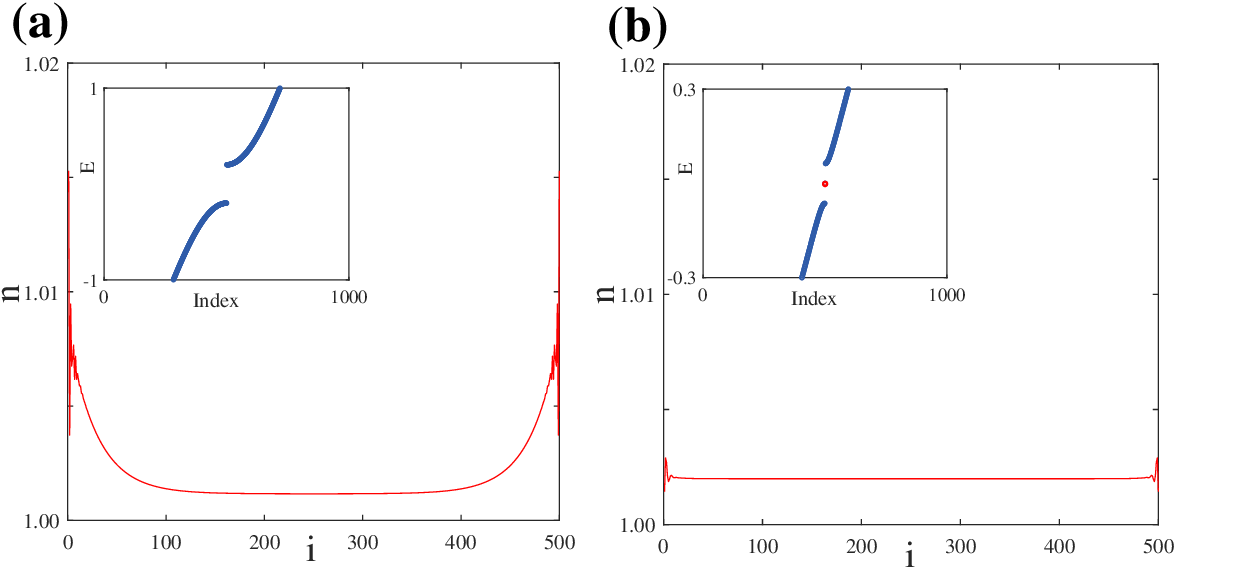}\caption{ (a) Particle distributions of the system. The upper plot illustrates the average quantity of particles per cell (number of cells is $L$, number of particles is $L+1$). Parameters are set to  $U=1.2$, $T=0.1$, $\gamma= J-\delta$. (b)  Particle distributions given   $U=0$, $T=0.15$, and $\gamma=- ( J-\delta)$. 
All figures are examined under parameters $L=500$,  $t=1$, $J^2=1.6\times 10^4$ and $\delta^2=2.5\times 10^{-10}$.} 
\label{Fig4}
\end{figure}

 Based on the above analysis, the four regions of the phase diagram correspond to whether the system state and the energy band structure are in a topological phase. Interestingly, besides the regions where both the bands and quantum states are either topological (region IV) or trivial (region I), there exist distinct cases: one where the quantum state is in   topological phase while the energy band structure is not (region II), and another with the energy band structure in  topological phase while the quantum state is not (region III). To visualize the verification, we calculated the state density distribution and the energy band for region II and region III in the phase diagram, as shown in Fig.\;\ref{Fig4}. In Fig.\;\ref{Fig4}(a), the system do not exhibit a zero mode, but calculation of the particle density shows accumulation near the edges, indicating the existence of edge states. In contrast, in Fig.\;\ref{Fig4}(b), while the system have a zero mode, there is no particle accumulation towards the edges---the slight particle density fluctuations at the boundary are a boundary effect which also present in topologically trivial Hermitian models and do not lead to an accumulation of particles at the edges. This result demonstrates the unique properties of the phase diagram of a many-body non-Hermitian topological phase: the band structure and quantum states can have different phase transition points.

 \section{State topological invariant} 

\begin{figure}[ptb]
\includegraphics[clip,width=0.5\textwidth]{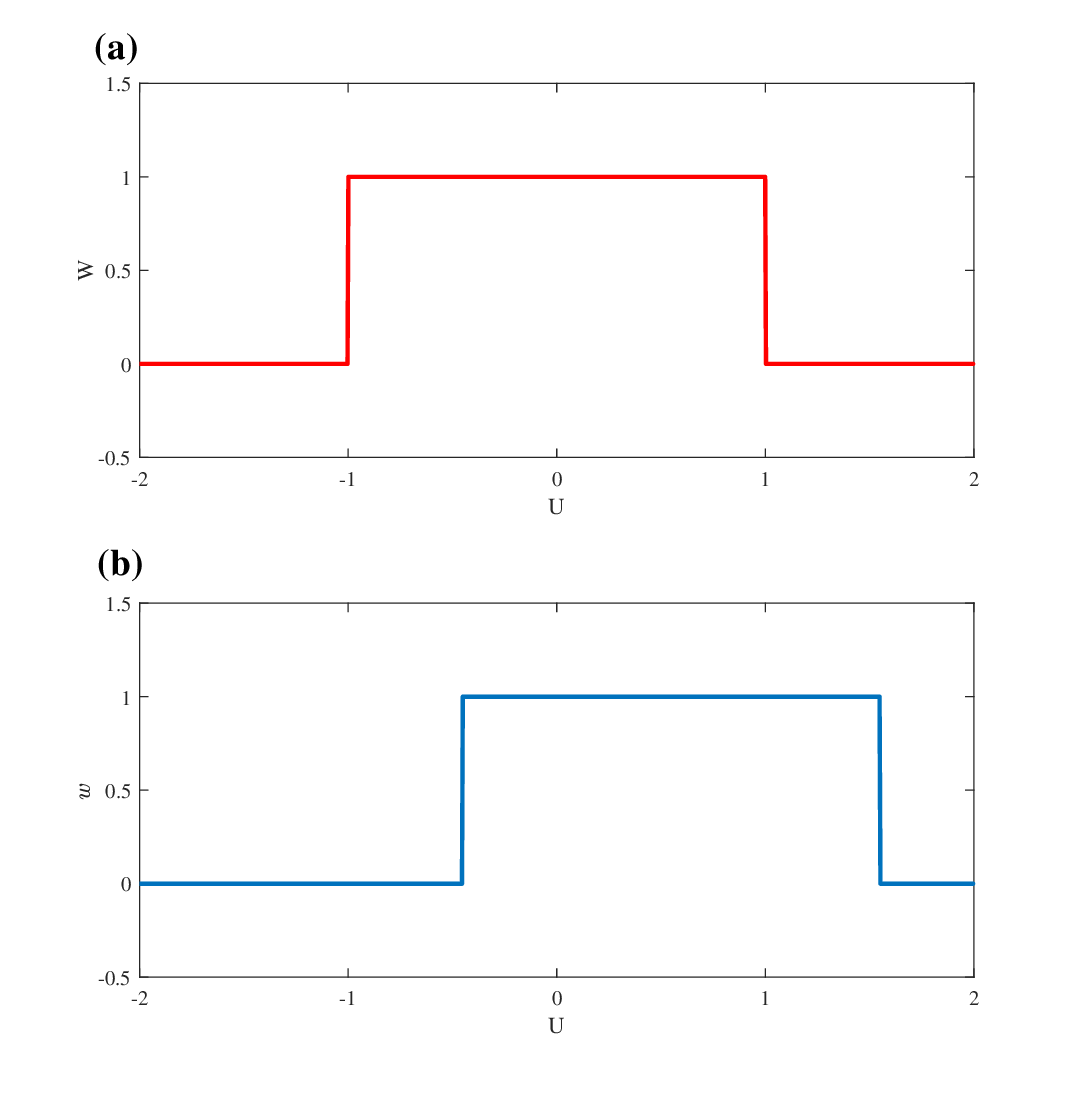}\caption{(a) Numerical result of the topological invariant $W$. (b) Numerical result of the topological invariant $w$.   We set $T=1$, $t=1$, $J=1$  and  $\gamma=0.5$.} 
\label{Fig5}
\end{figure}

 The above result implies that the many-body non-Hermitian systems need to be characterized by a pair of topological invariants $(W, w)$, which respectively describe the topology of the energy bands and the quantum states.
The topology of the energy bands is described by the topological invariant in the 38-fold classification. We present the topological invariant characterizing this model within the 38-fold classification. In this classification, the topological invariants are determined by the symmetry class of the system, which is dictated by time-reversal symmetry, particle-hole symmetry, chiral symmetry, and three related Hermitian conjugates, as well as the type of gap. 
  We consider the case where both $U$ and $t$ are non-zero. The model has chiral symmetry, time-reversal symmetry, particle-hole symmetry:
\begin{align}
 &\sigma_x {H_\mathrm{NH}^\dag}(k) \sigma_x= -{H_\mathrm{NH}}(k)  \nonumber   \\
 & {H_\mathrm{NH}^*}(k)={H_\mathrm{NH}}(-k)  \nonumber   \\
&  \sigma_x {H_\mathrm{NH}^T}(k) \sigma_x=-{H_\mathrm{NH}}(-k).
\end{align}
 The $\mathbb{Z}$ topological
invariant is well defined in the presence of a line gap, which is given as the winding number
\begin{equation}
W=\frac{1}{4\pi i} \int_{-\pi}^{\pi} dk  \; \mathrm{tr} \; \sigma_x H_\mathrm{NH}^{-1} (k)\frac{d}{dk} H_\mathrm{NH}(k).
\end{equation}
We have $W=1$ for $U_{g-}<U<U_{g+}$ and $W=0$ for $U>U_{g+}$ or $U<U_{g-}$, see Fig.\;\ref{Fig5}(a). The transition of the topological number $W$ signifies both the emergence of zero modes and the closure of the band gap. However, this topological invariant does not characterize the quantum states of the system.

We construct a topological invariant to characterize the topology of quantum states. Previous constructions use $H_{\mathrm{NH}}$ as the starting point, which should be revised to  reflect the properties of the many-body quantum state. Therefore, the topology of the state for the non-Hermitian system is characterized by the topological invariant of system with Hamiltonian $H_\mathrm{eff}$.
The effective model has chiral symmetry, time-reversal symmetry, particle-hole symmetry:
\begin{align}
 &\sigma_x {H_\mathrm{eff}^\dag}(k) \sigma_x= -{H_\mathrm{eff}}(k)  \nonumber   \\
 & {H_\mathrm{eff}^*}(k)={H_\mathrm{eff}}(-k)  \nonumber   \\
&  \sigma_x {H_\mathrm{eff}^T}(k) \sigma_x=-{H_\mathrm{eff}}(-k),
\end{align}
with topological invariant is defined as
\begin{equation}
w=\frac{1}{4\pi i} \int_{-\pi}^{\pi} dk  \; \mathrm{tr} \; \sigma_x H_\mathrm{eff}^{-1}(k) \frac{d}{dk} H_\mathrm{eff}(k).
\end{equation}

  By calculating the winding number, we find that its change does not occur at $U_{g\pm}$. Instead, the change occurs at $U_{c \pm}.$  We have $w=1$ for $U_{c-}<U<U_{c+}$ and $w=0$ for $U>U_{c+}$ or $U<U_{c-}$, see Fig.\;\ref{Fig5}(b).  The transition of the topological number, the emergence of boundary states, and the location of phase transitions without gap closing, all align consistently. Therefore, we have identified the correct topological invariant for the system.

In this model, the topological invariant $w$ is simply obtained by substituting $H_\mathrm{NH}$ in the topological invariant $W$ with $H_\mathrm{eff}$, but it is important to note that  in  general many-body non-Hermitian systems the state topological invariants are not defined simply by replacing the non-Hermitian Hamiltonian $H_\mathrm{NH}$ with its corresponding effective Hamiltonian $H_\mathrm{eff}$ in the topological invariants as defined within the framework of the 38-fold classification. One reason lies in the fact that the symmetries of the system's Hamiltonian generally do not correspond to the symmetries of the effective Hamiltonian. Consequently, the system may fall into different symmetry classes within the 38-fold classification, while the state remains within the same AZ symmetry class and is necessarily characterized by the same topological invariant. For example, for the model described by Eq.\;(\ref{EQ1}), when $U$ and $t$ are not all equal to zero, the system exhibits chiral symmetry, particle-hole symmetry, time-reversal symmetry. When $U=t=0$, the system has an additional sublattice symmetry  
\begin{equation}
 \sigma_z {H_\mathrm{NH}}(k) \sigma_z= -{H_\mathrm{NH}}(k).
\end{equation}
However,  the effective Hamiltonian $H_\mathrm{eff}(k) \propto A_{y}(k)\sigma_y+A_{z}(k)\sigma_z $, with
\begin{align}
A_{y}(k)  &  =  \sin k  \frac{ J^2-\gamma^2 }{J} ,\nonumber \\
A_{z}(k)  &  = -  \frac{\gamma}{2J} \frac{ \Delta_k}  
{\tanh \left(   \beta \Delta_k/2 \right)},
\end{align}
which yield for non-Hermitian case $\gamma \neq 0$
\begin{equation}
 \sigma_z H_\mathrm{eff}(k) \sigma_z \neq -{H_\mathrm{eff}}(k)
\end{equation}
do not have additional symmetry.  The symmetry of its effective Hamiltonian is  not increased, including the zero-temperature limit. The symmetry class of the effective Hamiltonian is always BDI. When breaking time-reversal symmetry and particle-hole symmetry by introducing certain non-reciprocity between cells, which changes  $\sin k$ to $\sin k + \delta$ in the bloch Hamiltonian, the symmetry of the effective Hamiltonian remains unaffected by the presence of sublattice symmetry at $U=t=0$, consistently belonging to class AIII.

Additionally, even if the system lacks certain AZ symmetries, the effective Hamiltonian of the system may still possess these AZ symmetries.  An example of this ineffectiveness can be considered the scenario where \begin{equation}\label{16}\hat{H}^{\prime}_{\mathrm{NH}}=(1-i)\hat{H}_{\mathrm{NH}}.\end{equation} Here $\hat{H}_{\mathrm{NH}}$  is defined in Eq.\;(\ref{EQ1}).  In this situation, the system will be trapped in the state with the largest imaginary part, which is the zero-temperature state of $\hat{H}_{\mathrm{NH}}$\;\cite{Kcaoprr}. That is, the  effective Hamiltonian $\hat{H}^{\prime}_{\mathrm{eff}}$, which corresponds to the non-Hermitian Hamiltonian $\hat{H}^{\prime}_{\mathrm{NH}}$ is $\hat{H}^{\prime}_{\mathrm{eff}}=\hat{H}_{\mathrm{eff}}(T\rightarrow 0)$, where  $\hat{H}_{\mathrm{eff}}$  is defined in Eq.\;(\ref{EQ7}).   The effective Hamiltonian still exhibits chiral symmetry $\sigma_x {H_\mathrm{eff}^{\prime}(k)^\dag} \sigma_x= -{H_\mathrm{eff}}^{\prime}(k)$, but the Hamiltonian of the system will change under the chiral operator.

The above discussion means that we require symmetry classes of non-Hermitian Hamiltonian  that is different from the 38 symmetric classes, and defining topological invariants that are associated with them.

\section{Symmetry class and topological invariants for non-Hermitian system's states} We deduce  the symmetry class and topological invariants that dictate the topology of quantum states in non-Hermitian systems. The completion of the symmetry class can be based on the following observation: while non-Hermitian systems have non-Hermitian Hamiltonians, their quantum states density matrix $\rho$ are invariably Hermitian, thus, it can be characterized by effective Hermitian Hamiltonian $\hat{H}_\mathrm{eff}\equiv -\ln \rho$ (apart from direct calculations, the effective Hamiltonian can be determined by the statistical mechanics of non-Hermitian systems, which described in  Appendix D). The effective Hamiltonian is classification according to the AZ symmetry class.  Consequently, the core issue in  obtaining the symmetry class of the Hamiltonian for non-Hermitian systems is to determine how the symmetry class of the effective Hamiltonian maps to that of the Hamiltonian of the non-Hermitian systems.  The AZ symmetry class is based on whether the system possesses particle-hole symmetry, chiral symmetry, see Table \ref{table:1}. \begin{table}[h!]
\centering
\begin{tabular}{ccccccc}
\hline\hline
Class & TRS &  PHS &
 CS  \\
\hline
$\mathrm{A}$ &   $0$  & $0$  & $0$  \\
$\mathrm{AI}$   & $+1$ & $0$ &  $0$  \\
$\mathrm{AII}$   & $-1$ & $0$ & $0$ \\
$\mathrm{AIII}$   & $0$ & $0$ & $1$ \\
$\mathrm{BDI}$   & $+1$ & $+1$ & $1$  \\
$\mathrm{CII}$   & $-1$ & $-1$ & $1$  \\
$\mathrm{D}$ & $0$ & $+1$ & $0$  \\
$\mathrm{C}$ & $0$ & $-1$ & $0$  \\
$\mathrm{DIII}$ & $-1$ & $+1$ & $1$  \\
$\mathrm{CI}$ & $+1$ & $-1$ & $1$ 
 \\
\hline\hline
\end{tabular}
\caption{AZ symmetry classes for Hermitian Hamiltonians.  TRS, PHS, CS respectively represent   time-reversal symmetry, particle-hole symmetry and  chiral symmetry.}
\label{table:1}
\end{table} In the table, $0$ signifies that the system's Hamiltonian lacks the corresponding symmetry, while $1$, $\pm1$ represents the presence of the same. Furthermore, $+1$ ($-1$) implies that the system not only embodies this symmetry, but the square of the respective first-quantized symmetry operator is equal to $+1$ ($-1$). Therefore, in order to understand how the symmetry class of the effective Hamiltonian maps to the symmetry class of the non-Hermitian Hamiltonian of non-Hermitian systems, we need to address three key points: 
(1) Identifying properties in the system's Hamiltonian correlating to particle-hole symmetry in the effective Hamiltonian and noting characteristics  when the square of the first-quantized particle-hole operator equals $+1$ or $-1$.
(2) Identifying the properties in the system's Hamiltonian that match with the chiral symmetry in the effective Hamiltonian. 
(3) Identifying properties in the system's Hamiltonian that correspond to time-reversal symmetry in the effective Hamiltonian and noting characteristics when the square of the first-quantized time-reversal operator equals $+1$ or $-1$.

 The following theorem provides the first step to solve this problem:

\textit{Theorem 1}: The Hamiltonian of a non-Hermitian system $\hat{H}_{\mathrm{NH}}$ and the system's  effective Hamiltonian $\hat{H}_\mathrm{eff}$ exhibit the following relation: $S_{\hat{G}}(\hat{H}_{\mathrm{NH}})=\hat{H}_{\mathrm{NH}}$  if and only if  $|\hat{G}  e^{-\hat{H}_\mathrm{eff}} \hat{G}^{\dag}|=|e^{-\hat{H}_\mathrm{eff}}|$ (for finite temperature situations, the system-environment coupling  also needs  to  invariant under the action of $S_{\hat{G}}$). Here the symbol $|\hat{A}|$ represents the operator norm  which defined as $|\hat{A}|=\frac{\hat{A}}{\mathrm{tr}(\hat{A})}$. $\hat{G}$ is an invertible operator such that $S_{\hat{G}}(\ ) \equiv \hat{G}( \ )\hat{G}^{-1}$ is linear under addition and satisfies $S_{\hat{G}}(\lambda \hat{A})=\lambda S_{\hat{G}}(\hat{A})$, where  $\lambda$ is a complex number and $\hat{A}$  is operator in the form of a free many-body system Hamiltonian.

We operate under the generally valid assumption that this steady state is unique. The proof of the theorem is provided in   Appendix B. This theorem can be interpreted as  the non-Hermitian Hamiltonian and quantum states share 
certain symmetry like $\hat{G}$.    
It illustrates that the particle-hole symmetry can be mapped from the  quantum state symmetry of the non-Hermitian system to the Hamiltonian's symmetry. We can take the logarithm on both sides of the equation satisfied by $\hat{H}_\mathrm{eff}$, resulting in $\hat{H}_{\mathrm{NH}}$ and $\hat{H}_\mathrm{eff}$ sharing the particle-hole symmetry.  For instance, for the model $\hat{H}_{\mathrm{NH}}$ described by Eq.\;(\ref{EQ1}),  the particle-hole symmetry  \begin{equation}\sigma_{x}H^T(k)\sigma_{x}= -H(-k)\end{equation} is  simultaneously preserved in non-Hermitian Hamiltonian and effective Hamiltonian.  The first key point has already been addressed.

  However, both time-reversal operator $\hat{T}$ and chiral operator $\hat{\Gamma}$ satisfy the condition of $S_{\hat{G}}(\lambda \hat{A})=\lambda^* S_{\hat{G}}(\hat{A})$ for quadratic operators $\hat{A}$ \cite{TCNH2}. Therefore, this theorem is not applicable to these symmetries.   An example of this ineffectiveness can be considered the scenario where $\hat{H}^{\prime}_{\mathrm{NH}}=(1-i)\hat{H}_{\mathrm{NH}}$. As discussed at the end of the previous section, the effective Hamiltonian still exhibits chiral symmetry, but the Hamiltonian of the system will change under the chiral operator. The cause of this ``symmetry breaking" resides in the fact that $\hat{T}$ and $\hat{\Gamma}$ typical transform gaining states into loss states. Therefore, time-reversal symmetry and chiral symmetry in the Hamiltonian represents a constraint between the many-body states with different eigenvalues, while the presence of above symmetry in the effective Hamiltonian implies certain constraints on the ground state itself. These are not the same requirement, hence the conclusion that time-reversal symmetry and chiral symmetry cannot be mapped between the effective Hamiltonian and the system Hamiltonian.

To take into account time-reversal symmetry and chiral symmetry, we define  the linearized time-reversal symmetry   and linearized chiral symmetry.

A system is said to possess linearized time-reversal symmetry (or linearized chiral symmetry) if  its Hamiltonian  satisfies 
\begin{equation}\label{EQ2}S_{\hat{G}^A_L} (\hat{H}_{\mathrm{NH}})=\hat{H}_{\mathrm{NH}},\end{equation}  with
\begin{equation*}
S_{\hat{G}^A_L} (\hat{H}_{\mathrm{NH}})\equiv (\hat{G}^A)_{ \{\left \vert n\right \rangle _{R} \}}  \hat{H}_{\mathrm{NH}}(\hat{G}^A)_{ \{\left \vert n\right \rangle _{R} \}}^{-1},\end{equation*}
 where  $\hat{G}^A$ is time-reversal operator $\hat{T}$ (or chiral operator $\hat{\Gamma}$),   $(\hat{G}^A)_{ \{\left \vert n\right \rangle _{R} \}} \equiv \sum_n (\hat{G}^A \left \vert n\right \rangle _{R} ) \left\langle n\right \vert _{L}$ with $\{\left \vert n\right \rangle _{R} \}$, $\{\left \vert n\right \rangle _{L} \}  $ is the set of biorthogonal left and right eigenstates of the Hamiltonian $\hat{H}_{\mathrm{NH}}$. 
 When acting on the effective Hamiltonian corresponding to the steady-state ensemble $\rho$ of   system with Hamiltonian $\hat{H}_{\mathrm{NH}}$,  we have 
\begin{equation}
\label{EQ3}
(\hat{G}^A)_{ \{\left \vert n\right \rangle _{R} \}} e^{-\hat{H}_\mathrm{eff}}  [\hat{G}^A)_{ \{\left \vert n\right \rangle _{R} \}}]^{\dag}=\hat{G}^A e^{-\hat{H}_\mathrm{eff}} (\hat{G}^A)^{-1}.
\end{equation}
Here, we utilize the anti-unitarity of $\hat{G}^A$ along with the fact that $e^{-\hat{H}_\mathrm{eff}}$ is a Hermitian matrix. Since $(\hat{G}^A)_{ \{\left \vert n\right \rangle _{R} \}}$ is a linear operator, the Theorem 1 can be applied. Combining on Eqs.\;(\ref{EQ2}) and (\ref{EQ3}), we obtain that:

\textit{Theorem 2}: The effective Hamiltonian of a non-Hermitian system has a  time-reversal symmetry (or  chiral symmetry) if and only if the system's Hamiltonian has linearized time-reversal symmetry (or linearized chiral symmetry).

We emphasize that in Theorem 2, the linearized time-reversal  symmetry (or linearized chiral symmetry) that the system's Hamiltonian satisfies specifically refers to the one constructed based on the time-reversal  symmetry (or  chiral symmetry) that the  system's effective Hamiltonian  satisfies.

The conditions for a single-body Hamiltonian to satisfy linearized time-reversal symmetry  and linearized chiral symmetry are (see details in Appendix C)
\begin{align}
\label{20}
 & T  {H}_{\mathrm{NH}}  T ^{-1} =  [\mathcal{C} ({H}_{\mathrm{NH}})]^*  \nonumber   \\
 & \Gamma  {H}_{\mathrm{NH}}  \Gamma^{-1} =  -[\mathcal{C} ({H}_{\mathrm{NH}})]^\dag,
\end{align}
where $T$, $\Gamma$ represent the part of unitary operators of first-quantized operator of ordinary time-reversal operator $\hat{T}$ and chiral operator $\hat{\Gamma}$, $\mathcal{C} ({H}_{\mathrm{NH}})$ is defined as the operator sharing the same eigenstates  
with ${H}_{\mathrm{NH}}$, but with eigenvalues corresponding to the complex conjugation of ${H}_{\mathrm{NH}}$'s eigenvalues.  When the Hamiltonian has a real spectrum, we can note that the linearized symmetry is equivalent to the original symmetry. This means that the system's Hamiltonian and its effective Hamiltonian share the same AZ symmetries under real spectrum scenarios, ensures consistency with the Hermitian case.  

The above discussion has already addressed the second key point---the chiral symmetry of the effective Hamiltonian corresponds to the linearized chiral symmetry of the system Hamiltonian.   In order to address the third key point, we define the square of the first-quantized linearized time-reversal operator $T_L^2$ as  $TT^*$. Combining with Theorem 2, it can be discerned that if the Hamiltonian of the non-Hermitian system possesses a linearized time-reversal symmetry, $T_L^2\equiv TT^*$ is also equated to the square of the first-quantized time-reversal operator pertinent to the time-reversal symmetry of the effective Hamiltonian. Therefore, we have addressed the third key point.

 Noteworthily, in the conventional topological classification of non-Hermitian systems, there exist certain extended symmetries, manifest in non-Hermitian Hamiltonians such as the sublattice
symmetry $\hat{S} \hat{H}\hat{S}^{-1}=\hat{H}^\dag$, which are not reflected in the effective Hamiltonian.  We would like to emphasize that the symmetry expressed as $\hat{S}\hat{H}\hat{S}^{-1}=\hat{H}^\dag$ will not impose constraints on the symmetry of the steady state.   An example has already been discussed at the end of the previous section. In this case, the effect of $\hat{S}$ is approximately that of $\hat{S} | m \rangle_{R} \sim| m \rangle_{L} $ ($| m \rangle_{R}$/$| m \rangle_{L}$ is the right/left many-body eigenstate), mapping the density matrix to $\rho \sim \sum_m P_m | m \rangle_{L} \langle m |_{L}$ that is inconsistent with the original density matrix. This result suggests that the extended symmetries in non-Hermitian systems are superfluous for the classification of quantum states. Although these symmetries can be reduced to AZ symmetries of quantum states in the Hermitian limit, safeguarding the presence of gapless states within the effective Hamiltonian. However, upon the introduction of non-Hermiticity, these AZ symmetries will be broken, leading to the gapping of the original gapless states. Consequently, the non-Hermitian systems with and without these extended symmetries are topologically indistinguishable from quantum state perspective.

Based on the above discussions, we have obtained the topological classification for non-Hermitian systems, see Table \ref{table:2}. In columns $2,3,4$, similar to Table \ref{table:1},  $0$ signifies that the system's Hamiltonian lacks the corresponding symmetry. The presence of $1$ or $\pm1$ indicates that the system possesses this symmetry. $+1$ ($-1$) denotes not only the existence of this symmetry in the system, but also that the square of the respective first-quantized symmetry operator equals $+1$ ($-1$).

\begin{table}[h!]
\centering
\begin{tabular}{ccccccc}
\hline\hline
Class & LTRS & PHS & LCS & $d=1$ & $d=2$ & $d=3$ \\
\hline
$\mathrm{A}$  & $0$ & $0$ & $0$ & $0$ & $Z$  & $0$ \\
$\mathrm{AI^*}$   & $+1$ & $0$ & $0$ & $0$ & $0$ & $0$ \\
$\mathrm{AII^*}$   & $-1$ & $0$ & $0$ & $0$ &$Z_2$ &$Z_2$ \\
$\mathrm{AIII^*}$   & $0$ & $0$ & $1$ & $Z$ & $0$ &$Z$ \\
$\mathrm{BDI^*}$   & $+1$ & $+1$ & $1$ & $Z$ & $0$ & $0$ \\
$\mathrm{CII^*}$   & $-1$ & $-1$ & $1$ & $Z$ & $0$ &$Z_2$ \\
$\mathrm{D}$ & $0$ & $+1$ & $0$ & $Z_2$ &$Z$ & $0$ \\
$\mathrm{C}$ & $0$ & $-1$ & $0$ & $0$ & $Z$ & $0$ \\
$\mathrm{DIII^*}$ & $-1$ & $+1$ & $1$ & $Z_2$ &$Z_2$ & $Z$ \\
$\mathrm{CI^*}$ & $+1$ & $-1$ & $1$ & $0$ & $0$ &$Z$ \\
\hline\hline
\end{tabular}
\caption{Classification of the quantum state of non-Hermitian topological insulators and superconductors in 1-3 spatial dimensions. LTRS, PHS, LCS respectively represent  linearized time-reversal symmetry, particle-hole symmetry and linearized chiral symmetry. }
\label{table:2}
\end{table}

The topological invariants for Table \ref{table:2}  can be derived. For classes $\mathrm{A},\mathrm{D},\mathrm{C}$ the topological invariant is  nothing but that defined in the AZ symmetry class, only the Hamiltonian needs to be replaced with the (single-body) effective Hamiltonian. For other classes that are marked as $\mathrm{X}^*$, the topological invariant is substituting the effective Hamiltonian into the topological invariant of the corresponding AZ symmetry class $\mathrm{X}$. If the topological invariant involve time-reversal and chiral operators, they are precisely the $T$ and $\Gamma$ defined in the expression of linearized time-reversal symmetry and linearized chiral symmetry, respectively.

For systems which Hamiltonians within a given symmetry class, quantum states with identical topological invariants can be continuously connected by tuning the Hamiltonian. Differing invariants necessitate a quantum phase transition, tied to the effective Hamiltonian’s gap closing, but not necessarily to that of the Hamiltonian itself. 

It's worth noting that there are systems with non-Hermitian skin effect \cite{Martinez,Yao2018,Yao20182,Lee2019,Kunst2018,Yin2018,   Slager2020, YYi, SMu2020, Roccati}. Effective Hamiltonian of those systems does not have translational symmetry under open boundary conditions---topology can not be well defined for such effective Hamiltonian\;\cite{ckprb}.  Therefore,  the quantum state's topology of the systems with significant skin effect not being well-defined.

\section{Conclusions} In this paper, we find that in many-body non-Hermitian systems, quantum state's topological phase transitions  can occur at points where the gap does not close. This indicates that the topology of many-body quantum states and energy bands should be considered independently, with previous works having focused solely on the topology of energy bands. We discover that the topology of states is described by topological invariant determined by the system Hamiltonian's particle-hole symmetry, linearized time-reversal symmetry, and linearized chiral symmetry.  Our research uncovers a distinctive facet of non-Hermitian topological systems, establishing a ground for the examination of quantum states in non-Hermitian topological insulators and superconductors---the framework put forth in this study enables explorations into quantum states in high-dimensional topological insulators and superconductors, paving the way for future progress within this research domain. Furthermore, our research methodology is extendable to the investigation of higher-order topological systems, weak topological systems, and floquet topological systems. Beyond this, our study also cracks open the door for delving into the phenomena of interacting many-body topological systems.

\acknowledgments This work was supported by the Natural Science Foundation of China (Grant No. 12174030) and National Key R$\&$D Program of China (Grant No. 2023YFA1406704). We are grateful to S.-Q. Zhao for helpful discussions.
 
\section*{ Appendix A: steady state Density matrix of  the non-Hermitian model}
\setcounter{equation}{0}
\renewcommand\theequation{A\arabic{equation}}
In our setup, the system comprises two parts: the non-Hermitian system and the heat bath. Therefore, the non-Hermitian system can be viewed as an open non-Hermitian system. Customarily, the time evolution of an open quantum system S is depicted by the quantum master equation. Following a derivation similar to that in the Hermitian case, we obtain \cite{BreuerPetruccione}
\begin{align}
\frac{d}{dt}\rho^{I}_{\mathrm{S}}(t)  &  =\sum_{a,b}\sum_{\omega}[\Gamma
_{ab}\left(  \omega \right)  (\hat{A}_{b}(\omega)\rho^{I}_{\mathrm{S}}%
(t)\hat{A}_{a}^{\dag}(\omega)\nonumber  \\
&  -\hat{A}_{a}(-\omega)\hat{A}_{b}(\omega)\rho^{I}_{\mathrm{S}}(t))+h.c.]. \label{10}
\end{align}
Here 
\begin{equation}
\label{7}\hat{A}_{a}(\omega)=\sum_{m}\left \vert m\right \rangle _{R}%
\left \langle m\right \vert _{L}\lambda_{a} \hat{C}_{a}\left \vert m+\omega
\right \rangle _{R}\left \langle m+\omega \right \vert _{L},
\end{equation}
with
\begin{align}
\Gamma_{ab}\left(  \omega \right)   &  =\int_{0}^{\infty}dte^{i\omega t}%
\mathrm{tr}_{\mathrm{B}}\left(  \hat{B}_{a}^{\dag}(t)\hat{B}_{b}(0)\rho
_{\mathrm{B}}^{I}\right)   
\end{align}
is the reservoir correlation function. In the above, $\left \vert m\right \rangle_{R}$ and $\left \vert m\right \rangle_{L}$ represent the bi-orthonormal right and left eigenstates of $\hat{H}_{\mathrm{NH}}$, respectively, both associated with eigenvalue $E_{m}$.  Further, $\left \vert m+\omega \right \rangle_{R/L} $ denotes the bi-orthonormal right/left eigenstate characterized by eigenvalue $E_{m}+\omega$. $a=1,2,...,2L$. $\lambda_a=\lambda^1_{a}$, $\hat{C}_{a}=a_{a}^{\dag}a_{a}$ for $a=1,2,...,L$; $\lambda_a=\lambda^2_{a-L}$, $\hat{C}_{a}=b_{a-L}^{\dag}b_{a-L}$ for $a=L+1,...,2L$.    $\rho
_{\mathrm{S}}^{I}$ and $\rho
_{\mathrm{B}}^{I}$ are the (unnormalized) density matrices  of the system and the thermal bath   in the interaction picture. 

Next, we  solve the time evolution equation of the non-Hermitian system to get the steady-state solution.  
Noticed that $\left \vert m\right \rangle _{R}/\left \vert m\right \rangle _{L}$
in Eq.\,(\ref{7}) can be expressed as%
\begin{equation}
\left \vert m\right \rangle _{R}=\mathcal{\hat{S}}\left \vert m\right \rangle
_{0}, \label{8}%
\end{equation}
\begin{equation}
\left \vert m\right \rangle _{L}=(\mathcal{\hat{S}}^{-1})^{\dag}\left \vert
m\right \rangle _{0}, \label{9}%
\end{equation}
where $\mathcal{\hat{S}}=  e^{\frac{1}{4}\ln  \frac{J+\gamma}{J -\gamma}  \sum_i   \psi^\dag_{i}   \sigma_z \psi_{i} } $ 
 and $\left \vert m\right \rangle _{0}$ is the eigenstate of a Hermitian Hamiltonian
$\hat{H}_{0}=\mathcal{\hat{S}}^{-1}\hat{H}_{\mathrm{HN}}\mathcal{\hat{S}%
} $. This expression can be verified to hold for models with both periodic boundary conditions and open boundary conditions.  According to $\hat{N}\left \vert
m\right \rangle _{R/L}=N\left \vert m\right \rangle _{R/L}$and $_{L}\left \langle
m|m\right \rangle _{R}=1$, $\left \vert m\right \rangle _{0}$ satisfy $\hat
{N}\left \vert m\right \rangle _{0}=N\left \vert m\right \rangle _{0}$ and
$_{0}\left \langle m|m\right \rangle _{0}=1$.

Substitute Eq.\,(\ref{8}), Eq.\,(\ref{9}) into Eq.\,(\ref{7}), and using Eq.\,(\ref{10}), we get
\begin{align}
\label{11}\frac{d}{dt}\rho^{I}_{\mathrm{S}}(t)  &  =\sum_{a,b}\sum_{\omega
}[\Gamma_{ab}\left(  \omega \right)  (\mathcal{\hat{S}}\hat{A}_{0,b}%
(\omega)\mathcal{\hat{S}}^{-1}\rho^{I}_{\mathrm{S}}(t)(\mathcal{\hat{S}}%
^{-1})^{\dag}\hat{A}_{0,a}^{\dag}(\omega)\mathcal{\hat{S}}^{\dag}\nonumber \\
&  -\mathcal{\hat{S}}\hat{A}_{0,a}(-\omega)\hat{A}_{0,b}(\omega)\mathcal{\hat
{S}}^{-1}\rho^{I}_{\mathrm{S}}(t))+h.c.],
\end{align}
where
\begin{equation*}
\hat{A}_{0,a}(\omega)=\sum_{m}\left \vert m\right \rangle _{0}\left \langle
m\right \vert _{0}\lambda_{a}\hat{C}_{a}\left \vert m+\omega \right \rangle
_{0}\left \langle m+\omega \right \vert _{0}.
\end{equation*}
Here $a,b=1,2,...,2L$.
Multiply both sides of the equal sign of Eq. (\ref{11}) by $\mathcal{\hat{S}%
}^{-1}$ to the left and $(\mathcal{\hat{S}}^{-1})^{\dag}$ to the right, we
get  
\begin{align}
&\frac{d}{dt}\mathcal{\hat{S}}^{-1}\rho^{I}_{\mathrm{S}}(t)(\mathcal{\hat{S}}^{-1})^{\dag}  \nonumber   \\ 
& =\sum_{a,b}\sum_{\omega}[\Gamma
_{ab}\left(  \omega \right)  (\hat{A}_{0,b}(\omega)\mathcal{\hat{S}}^{-1}%
\rho^{I}_{\mathrm{S}}(t)(\mathcal{\hat{S}}^{-1})^{\dag}\hat{A}_{0,a}^{\dag
}(\omega)  \nonumber  \\ 
& -\hat{A}_{0,a}(-\omega)\hat{A}_{0,b}(\omega)\mathcal{\hat{S}}^{-1}\rho^{I}_{\mathrm{S}}(t)(\mathcal{\hat{S}}^{-1})^{\dag}+h.c.]. \label{12}
\end{align}
Now, $\mathcal{\hat{S}}^{-1}\rho^{I}_{\mathrm{S}}(t)(\mathcal{\hat{S}}%
^{-1})^{\dag}$ obeys the master equation of the Hermitian system with Hamiltonian $\hat{H}_{\mathrm{0}}$. In the steady
state, the density matrix under the energy representation has only diagonal
terms. Using Eq.\,(\ref{12}), the diagonal terms of density matrix defined as
$P(n,t)=\left \langle n\right \vert _{0}\mathcal{\hat{S}}^{-1}\rho^{I}_{\mathrm{S}}(t)(\mathcal{\hat{S}}^{-1})^{\dag}\left \vert n\right \rangle
_{0}$ satisfy
\begin{equation}
\frac{d}{dt}P(n,t)=\sum_{m}[W(n|m)P(m,t)-W(m|n)P(n,t)] , \label{13}%
\end{equation}
where%
\begin{equation}
W(n|m)=\sum_{a,b}\gamma_{ab}(E_{m}-E_{n})\left \langle m\right \vert _{0}%
\lambda_{a}\hat{C}_{a}\left \vert n\right \rangle _{0}\left \langle
n\right \vert _{0}\lambda_{b}\hat{C}_{b}\left \vert m\right \rangle _{0}, 
\end{equation} 
with
\begin{align}
\gamma_{ab}(\omega) &  =\int_{-\infty}^{\infty}dte^{i\omega t} \mathrm{tr}_{\mathrm{B}}\left( \hat{B}_{a}^{\dag}(t)\hat{B}_{b}(0) \rho
_{\mathrm{B}}^{I} \right)  \nonumber  \\
&  \equiv \int_{-\infty}^{\infty}dte^{i\omega t}\left \langle \hat{B}_{a}^{\dag
}(t)\hat{B}_{b}(0)\right \rangle
\end{align}
is the real part of $2\Gamma_{ab}$. 

Using the Kubo-Martin-Schwinger  
condition $\left \langle \hat{B}_{a}^{\dag}(t)\hat{B}_{b}(0)\right \rangle
=\left \langle \hat{B}_{b}(0)\hat{B}_{a}^{\dag}(t+i\frac{1}{T})\right \rangle $,
we derive the temperature dependent behavior of $\gamma_{ab},$ i.e.,
\begin{equation}
\gamma_{ab}(-\omega)=e^{-\omega/ T}\gamma_{ba}(\omega). \label{14}
\end{equation}
When $\frac{d}{dt}P(n,t)=0$, Eq.\,(\ref{13}) and the relations Eq.\,(\ref{14})
give $W(n|m)e^{-\beta E_{m}}=W(m|n)e^{-\beta E_{n}}$ which lead to 
\begin{equation}
P(n)=const\times e^{-\beta E_{n}} 
\end{equation}
at the steady state. Then we have
\begin{equation}
\mathcal{\hat{S}}^{-1} \rho ^{I}_{T}(\mathcal{\hat{S}}%
^{-1})^{\dag}=\sum_{m}\left \vert m\right \rangle _{0}e^{-\beta E_{m}%
}\left \langle m\right \vert _{0} 
\end{equation}
or
\begin{equation}
\rho ^{I}_{T}=\sum_{m}\mathcal{\hat{S}}\left \vert m\right \rangle
_{0}e^{-\beta E_{m}}\left \langle m\right \vert _{0}\mathcal{\hat{S}}^{\dag},
\end{equation}
where $\rho ^{I}_{T}$ is the state at temperature $T$ in the interaction
picture, which is also the steady state
\begin{equation}
\rho_{T}=\sum_{m}\mathcal{\hat{S}}\left \vert m\right \rangle
_{0}e^{-\beta E_{m}}\left \langle m\right \vert _{0}\mathcal{\hat{S}}^{\dag
}%
\end{equation}
in the Schrödinger picture. Therefore, we have \begin{equation} \rho _{T}= \mathcal{\hat{S}}e^{-\beta \hat{H}_{0} }\mathcal{\hat{S}}^{\dagger}=e^{-\beta \hat{H}_\mathrm{NH}  } \mathcal{T}_c,\end{equation}
here  $\mathcal{T}_c=e^{\frac{1}{2}\ln  \frac{J+\gamma}{J -\gamma}  \sum_i   \psi^\dag_{i}   \sigma_z \psi_{i} }$.

\bigskip

\bigskip

\bigskip

\section*{ Appendix B: proof of Theorem 1}
\setcounter{equation}{0}
\renewcommand\theequation{B\arabic{equation}}
Firstly, we prove that if the effective Hamiltonian of a topological system satisfies $|\hat{G}  e^{-\hat{H}_\mathrm{eff}} \hat{G}^{\dag}|=|e^{-\hat{H}_\mathrm{eff}}|$,  the Hamiltonian of the system satisfies  $S_{\hat{G}}(\hat{H}_{\mathrm{NH}})=\hat{H}_{\mathrm{NH}}$. Here  $\hat{G}$ is an invertible operator such that $S_{\hat{G}}(\ ) \equiv \hat{G} (\ )\hat{G}^{-1}$ is linear under addition and satisfies $S_{\hat{G}}(\lambda \hat{A})=\lambda S_{\hat{G}}(\hat{A})$, where $\lambda$ is a complex number and $\hat{A}$  is operator in the form of a free many-body system Hamiltonian.

 In the Appendixes, we always treat the exceptional points case as a limiting case for systems without exceptional points, assuming that non-Hermitian Hamiltonians can always be diagonalized. We consider the Hilbert space to be finite-dimensional.

According to the definition of the effective Hamiltonian, $e^{-\hat{H}_\mathrm{eff}}$ is nothing but the density matrix of the formal system $\rho$. 
Without loss of generality, the system's steady-state density matrix can be written as:

\begin{equation}
\rho= \sum_m P_m | m \rangle_{R} \langle m |_{R}
\end{equation}
where  $| m \rangle_{R}$ is the  self-orthogonal right many-body eigenstate of the non-Hermitian Hamiltonian $\hat{H}_{\mathrm{NH}}$, $P_m$ is a real number  that represents the probability of the system being in state $| m \rangle_{R}$.

To ensure the stability of topological systems, we  assumption that small  changes in temperature do not affect the symmetry of the density matrix.  Such changes can  trigger a minor change in the occupancy of states. That is, we require that if $\rho_1= \sum_m P_{m} | m \rangle_{R} \langle m |_{R}$ is a density matrix possesses the symmetry $\hat{G}$ ($|\hat{G}  \rho_1 \hat{G}^{\dag}|=|\rho_1|$) then density matrix at another temperature $\rho_2=\sum_m P'_{m} | m \rangle_{R} \langle m |_{R}$  possesses the same symmetry $\hat{G}$ ($|\hat{G}  \rho_2 \hat{G}^{\dag}|=|\rho_2|$).

We define  $k_m=\frac{P_{m}}{ P'_{m}} $. We tune the temperature as much as possible to lift the degeneracy of $k_m$. However, there may still be some degeneracy that cannot be lifted under the above conditions, such as two energy levels $m$, $n$ with the same energy  in the Hermitian limit, where $k_m= k_n$ or $\frac{P_{m}}{ P_{n}}=1$  always holds.

 We remove duplicate $k$ and keep only one, retain only one $k$ from those that are degenerate. We renumber $k$, assuming that there are $N$ distinct $k$ left. We can define a function as:

\begin{equation}
\varepsilon(x) = \sum_{m=1}^{N} E_m \times \left( \prod_{{j=1 \atop j \neq m}}^{N} \frac{x - k_j }{k_m - k_j } \right).
\end{equation}

This function satisfies $\varepsilon(k_m) =E_m$.

Next, we   construct a operator $\hat{H}$ as
\begin{equation}
\hat{H}=  \varepsilon(\rho_1 \rho_2^{-1}).
\end{equation}
Due to $\rho_1 \rho_2^{-1} = \sum_m k_m | m \rangle_{R} \langle m |_{L}$, the operator $\hat{H}$ is 
$\hat{H} = \sum_m E_m | m \rangle_{R} \langle m |_{L}.$

We assert that  $\hat{H}_{\mathrm{NH}}$ belongs to $\hat{H}$.  The operator $\hat{H}_{\mathrm{NH}}$ does not belong to $\hat{H}$ if and only if there exist two energy levels $i, j$ such that $E_i \neq E_j$ and $k_i = k_j$, where the degeneracy indicated by $k_i = k_j$ cannot be lifted by adjusting the temperature. The existence of degeneracy $k_i = k_j$ that cannot be lifted by adjusting the temperature implies that there exist $i, j$ such that $\partial (\frac{P_i}{P_j})/ \partial T=0$  for some temperature region. However, after inspecting the statistical mechanics of non-Hermitian systems, we found that this case does not exist.   For the case in which $Im E_i=Im E_j$, non-Hermitian systems with stable states have $\frac{P_{i}}{ P_{j}}\propto e^{-  {Re (E_i-E_j)}/T }$. For $Im E_i$ not equal to $Im E_j$, $\frac{P_{i}}{ P_{j}}$ depends on the correlation functions of the heat bath and also on temperature $T$ \cite{Kcaoprr}.

We demonstrate the operator $\hat{H}$  has symmetry $\hat{G}$. Firstly, we prove that given $|\hat{G}\rho_1\hat{G}^{\dag}|=|\rho_1|$ and $|\hat{G}\rho_2\hat{G}^{\dag}|=|\rho_2|$,   $S_{\hat{G}}(\rho_1\rho_2^{-1})=\rho_1\rho_2^{-1}$. We have
\begin{equation}
 \rho_1\rho_2^{-1} \propto \hat{G}\rho_1\hat{G}^{\dag} (\hat{G}^{\dag})^{-1}\rho_2^{-1}\hat{G}^{-1} = S_{\hat{G}}(\rho_1\rho_2^{-1}).
\end{equation}
 Since $S_{\hat{G}}$ does not alter the trace of an operator, and the trace of $\rho_1\rho_2^{-1}$ is non-zero, we thus have $S_{\hat{G}}(\rho_1\rho_2^{-1})=\rho_1\rho_2^{-1}$.

Secondly, we prove that for each operator $\hat{A}$   in the form of a free many-body system Hamiltonian, $g[S_{\hat{G}}(\hat{A}) ]= S_{\hat{G}}[g(\hat{A})]$, here $g$ is analytic function. The functions of an operator are defined by Taylor series expansions.

To prove  this proposition, we use the Taylor expansion for $g$.  It can be represented as:
\begin{equation}
g(\hat{A}) = \sum_{n=0}^{\infty} \frac{g^{(n)}(a)}{n!} (\hat{A}-aI) ^n,
\end{equation}
where $g^{(n)}(a) $ is the $n$-th derivative of $g$ evaluated at point $a$, and $I$ is the identity operator.

Utilizing the properties of the operator $\hat{G}$, we have
\begin{align}
S_{\hat{G}}[g(\hat{A})] 
&= \sum_{n=0}^{\infty}  \frac{g^{(n)}(a)}{n!}\hat{G}(\hat{A} - aI)^n \hat{G}^{-1}  \nonumber \\
&= \sum_{n=0}^{\infty} \frac{g^{(n)}(a)}{n!} [\hat{G}(\hat{A} - aI)\hat{G}^{-1}]^n \nonumber \\
&= g[S_{\hat{G}}(\hat{A}) ].
\end{align}


Thirdly, we prove that if $\hat{\mathcal{A}}$ is a function of $\hat{A}$, then similarly we have $
g[S_{\hat{G}}(\hat{\mathcal{A}}) ]= S_{\hat{G}}[g(\hat{\mathcal{A}})]$.

Let us assume $\hat{\mathcal{A}}=f(\hat{A})$. Here $f$ is a function. Defining the function $l(x) = g[f(x)]$, we have 
\begin{align*}
S_{\hat{G}}[g(\hat{\mathcal{A}})] &= S_{\hat{G}}[gf(\hat{\mathcal{A}})] 
= l[S_{\hat{G}}(\hat{A}]  \nonumber \\ & = g f[S_{\hat{G}}(\hat{A})] =g[S_{\hat{G}}(\hat{\mathcal{A}}) ]= g[S_{\hat{G}}(\hat{\mathcal{A}}) ].
\end{align*}
Noting that $\rho_1 \rho_2^{-1}$ can always be regarded as a function of a free model many-body Hamiltonian $\hat{A}$, which has the same eigenstates as $\hat{H}_{\mathrm{NH}}$ but without energy degeneracy.  Therefore, by using $g[S_{\hat{G}}(\hat{\mathcal{A}}) ]= S_{\hat{G}}[g(\hat{\mathcal{A}})]$, we have

\begin{align}
S_{\hat{G}}(\hat{H}) &=S_{\hat{G}}[ \varepsilon(\rho_1 \rho_2^{-1})]
=\varepsilon (S_{\hat{G}}  (\rho_1 \rho_2^{-1}) ) 
=  \hat{H}.
\end{align}
We have proven that the Hamiltonian of the system possesses the $\hat{G}$ symmetry. 

Next,  we prove that if the  Hamiltonian of the system has  $S_{\hat{G}}(\hat{H}_{\mathrm{NH}})=\hat{H}_{\mathrm{NH}}$, and the system-environment coupling also invariant under the action of $S_{\hat{G}}$, the effective Hamiltonian of  the system satisfies $|\hat{G}  e^{-\hat{H}_\mathrm{eff}} \hat{G}^{\dag}|=|e^{-\hat{H}_\mathrm{eff}}|$. 

 We have ausmed the time evolution equation of the density matrix for a non-Hermitian system is given by  Eq.\;(\ref{1}). The Hamiltonian of the entire system comprises three components:
\begin{equation}
\hat{H}=\hat{H}_{\mathrm{NH}}\otimes \hat{I}_B+\hat{I}_S\otimes \hat{H}_{B}+\hat{H}_{BS},
\end{equation}
where $\hat{H}_{\mathrm{NH}}$ represents the non-Hermitian Hamiltonian of the system $S$, $\hat{H}_{B}$ corresponds to the Hamiltonian of the thermal bath, and $\hat{H}_{BS}$ denotes the coupling between the system and the thermal bath.  

The steady-state is given by
\begin{equation}
\label{29}
-i (\hat{H}\rho_s- \rho_s \hat{H}^{\dag}) =max(\lambda)  \rho_s.
\end{equation}  
Here $\lambda=(i[ \mathrm{tr} (\hat{H}^\dag-\hat{H}) \rho ] )$ is a real number, with $\frac{d}{dt}\rho=0$. We have $\mathrm{tr}_B \rho_s=|e^{-\hat{H}_\mathrm{eff}}|$. The steady state clearly satisfies $\frac{d}{dt}\rho_s=0$. However, states that satisfy $\frac{d}{dt}\rho=0$ are often not limited to the steady state alone.  We refer to these states as stationary states.  For   stationary states other than the steady state, they possess an $\lambda$ that is less than the $max(\lambda)$.

Firstly, we prove that, assuming  the  Hamiltonian of the system has  $S_{\hat{G}}(\hat{H}_{\mathrm{NH}})=\hat{H}_{\mathrm{NH}}$, and the system-environment coupling also invariant under the action of $S_{\hat{G}}$, if $\rho_s$ is the system's stationary state solution, i.e., $ \hat{H}\rho_s- \rho_s \hat{H}^{\dag} + [ \mathrm{tr} (\hat{H}^\dag-\hat{H}) \rho_s ] \rho_s=0$, then $\rho_s'=\frac{1}{Z}\hat{G}\rho_s \hat{G}^\dag$ is also the system's stationary state solution, where $Z=\mathrm{tr} \hat{G}\rho_s\hat{G}^\dag$. For the sake of notation simplicity, we have omitted the identity operator that is in direct product with $\hat{G}$. Substituting $\rho_s'=\frac{1}{Z}\hat{G} \rho_s\hat{G}^\dag$ into the time evolution equation, and using $\hat{G}\hat{H} \hat{G}^{-1}=\hat{H}  $ and $ \hat{H}\rho_s- \rho_s \hat{H}^{\dag} + [ \mathrm{tr} (\hat{H}^\dag-\hat{H}) \rho_s ] \rho_s=0$, we can obtain

\begin{align}
\label{30}
i \frac{d}{dt}\rho_s'= \frac{1}{Z} \hat{G} ( \hat{H}\rho_s- \rho_s \hat{H}^{\dag} + [ \mathrm{tr} (\hat{H}^\dag-\hat{H}) \rho_s ] \rho_s)\hat{G}^\dag=0.
\end{align}
Secondly, we prove that $\rho_s'$ is steady state. 
Using $\hat{H}_{\mathrm{NH}}= \hat{G}^{-1} \hat{H}_{\mathrm{NH}} \hat{G}$, 
$\frac{1}{Z} \hat{G} ( \hat{H}\rho_s- \rho_s \hat{H}^{\dag} + [ \mathrm{tr} (\hat{H}^\dag-\hat{H}) \rho_s ] \rho_s)\hat{G}^\dag=0$  can be rewritten as
$-i (\hat{H}\rho_s'- \rho_s' \hat{H}^{\dag}) =max(\lambda)  \rho_s'$. Therefore, we have shown that $\rho_s'$ is a steady state. As we have assumed that the steady state is unique,   we have $\rho_s'=\rho_s$ or $|\hat{G}  e^{-\hat{H}_\mathrm{eff}} \hat{G}^{\dag}|=|e^{-\hat{H}_\mathrm{eff}}|$.

It will be shown later that the  the coupling between the system and the environment does not affect the density matrix in the zero-temperature limit. Therefore, at zero temperature, there is no need for the condition that the system-environment coupling remains invariant under the action of $S_{\hat{G}}$.

Now, we have proved the Theorem 1 in the main text.

For operators that do not satisfy the condition $S_{\hat{G}}(\lambda \hat{A})=\lambda S_{\hat{G}}(\hat{A})$, such as time-reversal  operator, although Eq. (\ref{30}) holds and $\rho_s'$ remains a steady state, Eq. (\ref{29}) becomes $-i (\hat{H}\rho_s'- \rho_s' \hat{H}^{\dag}) =[-max(\lambda)]  \rho_s'$, indicating that this state is not a steady state, which can not 
lead to   $|\hat{G}  e^{-\hat{H}_\mathrm{eff}} \hat{G}^{\dag}|=|e^{-\hat{H}_\mathrm{eff}}|$.

\bigskip

\section*{ Appendix C: proof of Eq.\;(\ref{20})}
\setcounter{equation}{0}
\renewcommand\theequation{C\arabic{equation}}

First, we derive the first line of Eq.\;(\ref{20}). Expanding the Hamiltonian of the system in its biorthonormal eigenstates and utilizing the definition of $\hat{T}_L$, we obtain
\begin{align}
\label{C1}
& \hat{T}_L   \hat{H}_\mathrm{NH} (\hat{T}_L)^{-1}   \nonumber \\
&=   \sum_n (\hat{T} \left \vert n\right \rangle _{R} ) \left\langle n\right \vert _{L}  \sum_m E_m | m \rangle_{R} \langle m |_{L} \sum_k  \left \vert k\right \rangle _{R}   (\left\langle k\right \vert _{L} \hat{T}^{-1} ) \nonumber \\
 & =    \sum_n \sum_m \sum_k E_m (\hat{T} \left \vert n\right \rangle _{R} )      \delta_{mn} \delta_{mk} (\left\langle k\right \vert _{L} \hat{T}^{-1})  \nonumber \\
&= \sum_m E_m  \hat{T} | m \rangle_{R} \langle m |_{L} \hat{T}^{-1}  \nonumber \\
&=  \hat{T}( \sum_m   E_m^*  | m \rangle_{R} \langle m |_{L}) \hat{T}^{-1}  \nonumber \\
&= \hat{T}  \mathcal{C}(\hat{H}_\mathrm{NH}) \hat{T}^{-1}.
\end{align}


Therefore,   $\hat{T}_L   \hat{H}_\mathrm{NH} (\hat{T}_L)^{-1}  = \hat{H}_\mathrm{NH}$ is  equivalent to $\hat{T}  \mathcal{C}(\hat{H}_\mathrm{NH}) \hat{T}^{-1}=\hat{H}_\mathrm{NH}$.

By comparing the eigenvalues and eigenstates, we can observe that $\mathcal{C}(\hat{H}_\mathrm{NH})= \psi^\dag \mathcal{C}(H_\mathrm{NH}) \psi$. Here $\psi=(\psi_1,\psi_2,\psi_3,...,\psi_n)$, $(\psi_i)_{i=1,2,...n}$  is a set
of fermion annihilation operators for a normal system or
Nambu spinors for a superconductor, $\hat{H}_{\mathrm{NH}}= \psi^\dag {H}_{\mathrm{NH}}\psi $.
Utilizing \begin{equation}\hat{T}  \hat{H}  \hat{T}^{-1}=\psi^\dag [T^{-1}   {H} ^* T] \psi, \end{equation}
where  $T$ is the part of unitary operators of  first-quantized  $\hat{T}$, defined by  $\hat{T} \psi \hat{T} ^{-1}= T \psi$, we can obtain that
\begin{equation}
 \psi^\dag  {H}_\mathrm{NH}   \psi=  \psi^\dag T^{-1}  [\mathcal{C}({H} )]^* T\psi 
\end{equation}
or 
\begin{equation}
 T  {H}_{\mathrm{NH}}  T ^{-1} =  [\mathcal{C} ({H}_{\mathrm{NH}})]^*.
\end{equation}
  For the second line of Eq.\;(\ref{20}), imitating Eq.\;(\ref{C1}), it can be proved that $\hat{\Gamma}_L   \hat{H}_\mathrm{NH} (\hat{\Gamma}_L)^{-1} = \hat{\Gamma}  \mathcal{C} (\hat{H}_\mathrm{NH}) \hat{\Gamma} ^{-1}$, where $\hat{\Gamma}$ is the chiral operator. Utilizing $\hat{\Gamma}  \hat{H}  \hat{\Gamma}^{-1}=-\psi^\dag [\Gamma^{-1}   {H} ^\dag \Gamma] \psi $, where  $\Gamma$ is the part of unitary operators of  first-quantized  $\hat{\Gamma}$, we have

\begin{equation}\Gamma  {H}_{\mathrm{NH}}  \Gamma^{-1} =  -[\mathcal{C} ({H}_{\mathrm{NH}})]^\dag .\end{equation}

\section*{ Appendix D: Effective Hamiltonian of non-Hermitian topological insulator and topological superconductor}
\setcounter{equation}{0}
\renewcommand\theequation{D\arabic{equation}}
In this part, we present the effective Hamiltonian of non-Hermitian topological insulator and topological superconductor. Following the Hermitian case, we define the quantum states of non-Hermitian systems as the half-filling steady-state evolving under the dynamics of many-body non-Hermitian systems. We consider the finite temperature scenario, the zero-temperature scenario can be obtained as the limit of temperature approaching zero. The finite temperature non-Hermitian system is defined as a non-Hermitian systems weakly interacting with Hermitian thermal baths, as this approach is better aligned with the practical aspects of experimental setups. The time evolution equation of the density matrix for a non-Hermitian system is given by Eq.\;(\ref{1}).

 It is important to note that, under this definition, an imaginary number can be added to the non-Hermitian Hamiltonian without affecting the time evolution equation of the density matrix. Therefore, for simplicity, we always adjust the maximum imaginary part of the eigenvalue of the non-Hermitian system to zero. The coupling term, $\hat{H}_{BS}$, can be generally expressed as:
\begin{equation}
\hat{H}_{BS} = \sum_{a} \lambda_{a} \hat{C}_{a} \otimes \hat{B}_{a} + h.c.,
\end{equation}
where $\lambda_{a}$ is the real coupling strength, $\hat{C}_{a}$ is the operator acting on the system, and $\hat{B}_{a}$ is the operator acting on the thermal bath, $a=1,2,3,...,n$. There are often multiple expressions for the same coupling. We choose the operators $\hat{C}_{a}, \hat{C}^{\dag}_{a}$ and $\hat{B}_{a}, \hat{B}^{\dag}_{a}, \hat{I}_B$ to be linearly independent.     

The steady-state of such systems was investigated in Ref \cite{Kcaoprr}. The results showed that stability of such non-Hermitian system (for an unstable system, its steady state is determined by all the details of the heat bath and the system-environment coupling, and there may not even be a steady state at all) necessitates the existence of a single path-dependent conserved quantity $P_c(t)\mathcal{T}_c$, where $P_c(t) \equiv  \exp(2\int_{0}^{t} dt \left \langle \hat{\varUpsilon}\right \rangle 
) $ \ is a path-dependence factor.  $\left \langle \hat{\varUpsilon}\right \rangle \  $is
the expectation value of $\hat
{\varUpsilon}$ for non-Hermitian system's density matrix $\rho_S(t)$ at time $t$, and $\hat
{\varUpsilon}$ is defined as the non-Hermitian part of non-Hermitian Hamiltonian, i.e., $  \hat
{\varUpsilon} \equiv  \frac{1}{2i}(\hat{H}_{\mathrm{NH}} 
-\hat{H}_{\mathrm{NH}}^{\dagger}) $.  $\mathcal{T}_c$ is a  positive defined Hermitian operator in the non-Hermitian system.
The equilibrium state of thermalizable  quasi-Hermitian systems [with conserved quantity $P_c(t)\mathcal{T}_c$, up to a normalization factor] at temperature $T$ is \begin{equation}\rho_{\mathrm{NH}}=e^{-\beta_{T}\hat{H}_{\mathrm{NH}}%
}\mathcal{T}_c .\end{equation}
It is essential to note that the Boltzmann distribution, constructed under the assumption that the probability of states with energy $E_{n}$ is $P_{n} \propto e^{-\beta_{T}E_{n}}$, is generally not satisfied. The probability of states with energy $E_{n}$ is $P^\mathrm{NH}_{n} \propto e^{-\beta_{T}E_{n}}W_n \neq e^{-\beta_{T}E_{n}}$, where $W_n = \langle E_n |_{L} \mathcal{T}_c | E_n \rangle_{L}$, with $ | E_n \rangle_{L}$ is the left-eigenstate   which biorthogonal to  (self-normalized) $\left \vert E_n \right \rangle _{R}$.

For thermalizable  non-Hermitian systems, two sets of data, namely the system's Hamiltonian $\hat{H}_{\mathrm{NH}}$ and the set of operators $\{\hat{C}_a\}$ in the system-environment coupling, can uniquely determine $\mathcal{T}_c$ (up to an irrelevant multiplicative factor that leaves the expectation values of physical quantities unaffected), subject to the following two conditions:

(1A) Symmetric condition: the coupling operator $\hat{C}_{a}$ satisfies $[\hat{C}_{a},\mathcal{T}_c]=0$ for all $a$. Alternatively, in physical terms, the coupling needs to have symmetry $\mathcal{T}_c$.

(2A) Conjugacy relation: $\hat{H}_{\mathrm{NH}}\mathcal{T}_c-\mathcal{T}_c\hat{H}_{\mathrm{NH}}^{\dagger}=0$.

Systems that strictly possess the path-dependence conserved quantity are necessarily real-spectrum systems. For non-Hermitian Hamiltonians possessing complex eigenvalues,  while the system initially lacks exact conserved quantities, the eigenstates with negative imaginary parts will loss (we have adjust the maximum imaginary part of the eigenvalue of the non-Hermitian system to zero) and lead to the rapid appearance of an approximate conserved quantity $P_c^R(t)\mathcal{T}_c^{R}$. During the evolution of time, in the weak coupling limit, these states with negative imaginary part will decay rapidly on timescale $\tau_{loss}$ (compared to the timescale of non-loss state evolution $\tau_{nonloss}$), and hence do not significantly impact the dynamics of non-loss states. In this case, the system can be approximately regarded as a real-spectrum system, which has an error of approximately $\tau_{nonloss}/\tau_{loss}$. Consequently, the steady state of the system can be well described by the Hamiltonian restricted to the subspace of non-dissipative states.  
Therefore, in the case of a non-Hermitian system, the original  model is reduced to a subspace consisting of degenerate quantum states that share the same maximum imaginary part among all eigenvalues. Specifically, the Hamiltonian of the system and the coupling between the system and the bath restricted to this subspace  are:
\begin{equation}
\hat{H}^R_{\mathrm{NH}}=P\hat{H}_{\mathrm{NH}}P,
\end{equation}
\begin{equation}
\hat{H}_{BS}^{R}=\sum_{a} \lambda_{a} [( P \hat{C}_{a} P )\otimes \hat{B}_{a}+( P \hat{C}^\dag_{a} P)\otimes \hat{B}^\dag_{a}] .
\end{equation}
Where $P=\sum_{Im E_{m} = Max (Im E_{n})}\  \left \vert m\right \rangle _{R}\left \langle
m\right \vert _{L}$ is the projection operator to the subspace with maximum
imaginary part.

The steady state of the system (ignore the  corrections from lossy states to non-lossy states) is given by the following formula:
\begin{equation}\rho_{\mathrm{NH}}=e^{-\beta_{T}%
 \hat{H}^R_{\mathrm{NH}} }\mathcal{T}_c^{R}.\end{equation}

$\mathcal{T}_c^R$ is uniquely determined by the following relation (for thermalizable non-Hermitian systems):   

 (1B) The coupling operator $\hat{C}_{a}$ satisfies $ P\hat{C}_{a}P \mathcal{T}_c^{R}-\mathcal{T}_c^{R} P^\dag \hat{C}_{a}^\dag P^\dag=0$ for all $a$.  

(2B)   $\hat{H}^R_{\mathrm{NH}}\mathcal{T}_c^{R}-\mathcal{T}_c^{R}(\hat{H}^R_{\mathrm{NH}})^{\dagger}=0$.

\subsection{ Real spectrum case}
 Since here focuses on topological insulator or topological superconductor, $\hat{H}_{\mathrm{NH}}$ can be expressed as 
$\hat{H}_{\mathrm{NH}}= \psi^\dag {H}_{\mathrm{NH}}\psi $
where the matrix ${H}_{\mathrm{NH}}$ is a single-body
non-Hermitian Hamiltonian.  $\psi=(\psi_1,\psi_2,\psi_3,...,\psi_n)$, $(\psi_i)_{i=1,2,...n}$  is a set
of fermion annihilation operators for a normal system or
Nambu spinors for a superconductor.

For the above Hamiltonian, $\mathcal{T}_c$ can be written as 

\begin{equation}\mathcal{T}_c=  \exp (\psi^\dag \ln T_c \psi) .
\end{equation}
 Here $T_c$ satisfying the condition 

(1C) The coupling operator $\hat{C}_{a}$ satisfies $ [\psi^\dag \ln T_c \psi,\hat{C}_{a}]=0$  for all $a$.

(2C) ${H}_{\mathrm{NH}}T_c  -  T_c {H}_{\mathrm{NH}}=0$.

Therefore, single-body effective Hamiltonian $ H_\mathrm{eff}$ defined as $e^{ - \psi^{\dag}  H_\mathrm{eff} \psi } =\rho$ can be written as
$H_\mathrm{eff}= - \ln (e^{-\beta {H}_{\mathrm{NH}}} T_c)$.  The corresponding many-body Hamiltonian is
$\hat{H}_\mathrm{eff}= - \psi^\dag \ln e^{-\beta {H}_{\mathrm{NH}}} T_c \psi$.

Base on $P^\mathrm{NH}_{n} \propto e^{-\beta_{T}E_{n}}W_n$, for the zero-temperature case, the system is always in the state of lowest energy for all $T_c$ meet condition (2C). At zero temperature, the coupling between the system and environment does not affect the steady state of a non-Hermitian system. Therefore, we can arbitrarily choose $T_c$ that satisfies condition (2C) to calculate the effective Hamiltonian at zero temperature (as   temperature tends to zero, the effective Hamiltonian often diverges. We can introduce a multiplicative factor that keeps the matrix elements of the effective Hamiltonian from diverging without altering the topology).

\subsection{Complex sepctrum case}
According to $\rho_{\mathrm{NH}}=e^{-\beta_{T}%
 \hat{H}^R_{\mathrm{NH}} }\mathcal{T}_c^{R}$, the probability of states $| E_n \rangle_{R}$ with energy $E_{n}$ is $P^\mathrm{NH}_{n}= \frac{1}{Z} B^R_n W^R_n  $, where 
\begin{equation}
B^R_n=
\begin{cases}
e^{-\beta_{T}Re E_{n}} & Im E_{n}= Max (Im E_{n}) \\
0    & Im E_{n} \neq Max (Im E_{n})
\end{cases}
,
\end{equation}
and $W^R_n = \langle E_n |_{L} \mathcal{T}^R_c | E_n \rangle_{L}$, with $ | E_n \rangle_{L}$ is the left-eigenstate   which biorthogonal to  self-normalized $\left \vert E_n \right \rangle _{R}$. Because these eigenstates and eigenvalues are associated with the many-body Hamiltonian, it makes   difficult to directly solve for the effective Hamiltonian.
To address this difficulty, we provide a   theorem:

Theorem 3: The steady state of system with  Hamiltonian $\hat{H}_{\alpha \rightarrow \infty}$ and coupling  $\{ \hat{C}_a \}$ is same as the steady state of  system with  Hamiltonian $\hat{H}_{\mathrm{NH}}$ and same coupling $\{ \hat{C}_a \}$. In the above, $\hat{H}_{\alpha  }$ is defined as $\hat{H}_{\alpha  }\equiv \psi^\dag {H}_{\alpha  } \psi $, where \begin{equation}    H_{\alpha  }=    Re H_{\mathrm{NH}} - \alpha Im H_{\mathrm{NH}}.\end{equation} 
 $Re {H}_{\mathrm{NH}}$ is defined as sharing the same eigenstates  
with ${H}_{\mathrm{NH}}$, but with eigenvalues corresponding to the real  
part of ${H}_{\mathrm{NH}}$'s eigenvalues, i.e., if ${H}_{\mathrm{NH}}=\sum_m E_m | m \rangle_{R} \langle m |_{L}   $, $Re {H}_{\mathrm{NH}} =\sum_m Re(E_m) | m \rangle_{R} \langle m |_{L} $.  $Im {H}_{\mathrm{NH}}$ is defined as sharing the same eigenstates  
with ${H}_{\mathrm{NH}}$, but with eigenvalues corresponding to the real  
part of ${H}_{\mathrm{NH}}$'s eigenvalues, i.e., if ${H}_{\mathrm{NH}}=\sum_m E_m | m \rangle_{R} \langle m |_{L}   $, $Im {H}_{\mathrm{NH}} =\sum_m Im(E_m) | m \rangle_{R} \langle m |_{L} $. 

  This theorem implies that the steady states of a non-interacting non-Hermitian system with a complex energy spectrum can be transformed into solving for the steady states of a non-interacting real spectrum system  with single-body Hamiltonian $H_{\alpha \rightarrow \infty}$, and then solved using the method introduced in subsection $A$.

\textit{Proof}.
Note that $\hat{H}_{\alpha  }\equiv \psi^\dag {H}_{\alpha  } \psi $ and $\hat{H}_{\mathrm{NH}}$ have the same set of eigenstates. Therefore, to prove that steady state of $\hat{H}_{\alpha  }$ is also the steady state of $\hat{H}_{\mathrm{NH}}$, we only need to compare the probability distributions of the eigenstates.
For system with Hamiltonian $H_\alpha \rightarrow \infty$,
 the probability of the eigenstate   $| E_n \rangle_{R}$ with energy $E_n$ is $P^\mathrm{NH}_{n} =\frac{1}{Z} B_n W_n  $
where $B_n=e^{-\beta_{T}Re E^{\alpha \rightarrow \infty}_{n}}$ and $W_n = \langle E_n |_{L} \mathcal{T}_c | E_n \rangle_{L}$.

First, we verify that the  factors $B_n$ of the two systems are the same. 

We have $B_n  =\frac{1}{Z} e^{-\beta_{T}   Re E_{n}} e^{\beta_{T} \alpha  [lm E_{n}-Max(lm E_{n})] }$ ($\alpha \rightarrow \infty$), or
\begin{equation}
B_n=
\begin{cases}
e^{-\beta_{T}Re E_{n}} & Im E_{n}= Max (Im E_{n}) \\
0    & Im E_{n} \neq Max (Im E_{n})
\end{cases}
,
\end{equation}

which is same as $B^R_n$.

Due to $B_n$ for the state $Im E_{n} \neq Max (Im E_{n})$ equals 0,  the condition $P^R_n = P_n$ only necessary to demonstrate that for states with $Im E_{n}= Max (Im E_{n})$,  $W^R_n = W_n$.
Next, we verify that the for for states with $Im E_{n}= Max (Im E_{n})$, $W^R_n = W_n$ is the same.

 We prove that $\mathcal{T}_c^{{R}}= P \mathcal{T}_c P^\dag$ (up to an irrelevant multiplicative factor).  Since  $\mathcal{T}_c^{{R}}$ is unique up to an irrelevant multiplicative factor, we only need  to prove that $P \mathcal{T}_c P^\dag$ satisfies conditions (1B) and (2B).
 
According to  condition (2A), $\mathcal{T}_{c}$ can be expressed as:
\begin{equation}
\mathcal{T}_{c}=\sum_{m}\  t_{m} \left \vert m\right \rangle _{R}\left \langle
m\right \vert _{R},
\end{equation}
 $\left \vert m\right \rangle _{R}/\left \vert m\right \rangle _{L}$ represent the biorthogonality right/left eigenstates of $\hat{H}_{\mathrm{NH}}$. $\{ t_{m} \}$ is   a set of positive real number. By utilizing the definition of $P$, we can verify that $P\mathcal{T}_c=\sum_{m,
E_m=Max (Im E_{n})}\  t_{m} \left \vert m\right \rangle _{R}\left \langle
m\right \vert _{R}= \mathcal{T}_c P^\dag$. 

 We prove that $\mathcal{T}_c^{{R}}= P \mathcal{T}_c P^\dag$ satisfies condition (1B). We have
\begin{align}
& P\hat{C}_{a}P (P \mathcal{T}_c P^\dag) -(P \mathcal{T}_c P^\dag) P^\dag \hat{C}_{a}P^\dag \nonumber \\ &  =   P\hat{C}_{a}P \mathcal{T}_c P^\dag -P \mathcal{T}_c   P^\dag \hat{C}_{a}P^\dag\nonumber \\
& = P\hat{C}_{a}  \mathcal{T}_c P^\dag -P \mathcal{T}_c    \hat{C}_{a}P^\dag  \nonumber \\
& =P[\hat{C}_{a},\mathcal{T}_c]P^\dag=0.
\end{align}
  We have used that $P^2=P$.

According to $\hat{H}^R_{\mathrm{NH}}=P\hat{H}_{\mathrm{NH}}P$, we verify that $\mathcal{T}_c^{{R}}= P \mathcal{T}_c P^\dag$ satisfies condition (2B):
 \begin{align}
&\hat{H}^R_{\mathrm{NH}} P \mathcal{T}_c P^\dag  - P \mathcal{T}_c P^\dag P^\dag (\hat{H}^R_{\mathrm{NH}})^\dag \nonumber \\ &=P\hat{H}_{\mathrm{NH}}P (P \mathcal{T}_c P^\dag) -(P \mathcal{T}_c P^\dag)P^\dag\hat{H}_{\mathrm{NH}}^\dag P^\dag\nonumber \\ &  =   P\hat{H}_{\mathrm{NH}}P \mathcal{T}_c P^\dag -P \mathcal{T}_c   P^\dag \hat{H}_{\mathrm{NH}}^\dag P^\dag\nonumber \\
& = P\hat{H}_{\mathrm{NH}}  \mathcal{T}_c P^\dag -P \mathcal{T}_c    \hat{H}_{\mathrm{NH}}^\dag P^\dag  \nonumber \\
& =P(\hat{H}_{\mathrm{NH}} \mathcal{T}_c-\mathcal{T}_c    \hat{H}_{\mathrm{NH}}^\dag)P^\dag=0.
\end{align}
Therefore, $P \mathcal{T}_c P^\dag$ meet the condition (1B) and (2B),  and it is the $\mathcal{T}_c^{{R}}$ of the system with Hamiltonian $\hat{H}_{\mathrm{NH}}$ and coupling  $\{ \hat{C}_a \}$.

For state $| E_n \rangle_{L}$ with $Im E_{n}= Max (Im E_{n})$, we have $P^\dag| E_n \rangle_{L}=| E_n \rangle_{L}$. Therefore 
  \begin{align}W_n=\langle E_n |_{L} \mathcal{T}_c | E_n \rangle_{L}= \langle E_n |_{L}  P\mathcal{T}_c P^\dag | E_n \rangle_{L} \nonumber \\ = \langle E_n |_{L} \mathcal{T}^R_c | E_n \rangle_{L}=W_n^R   \end{align}
 for $Im E_{n}= Max (Im E_{n})$.
 
Therefore, we prove that the two systems have the same steady state.


\begin{thebibliography}{50}
\bibitem {note} The impact of self-energy terms on the topology of the system has already been discussed in Ref.\;\cite{ZNTP}.
\end{thebibliography}
\end{document}